\documentclass[12pt]{elsarticle}
\usepackage{microtype}
\usepackage{slashed}
\usepackage[a4paper,textwidth=15.5cm]{geometry}
\usepackage{amsmath}
\usepackage{amssymb}
\usepackage[breaklinks=false]{hyperref}
\usepackage{xcolor}
\usepackage{listings}
\usepackage{booktabs}
\usepackage{tabularx}
\usepackage{graphics}
\graphicspath{{figures/}}

\fontencoding{T1}\selectfont

\DeclareMathOperator{\im}{Im}

\newcommand{\MS}{\ensuremath{\overline{\text{MS}}}}

\newenvironment{absolutelynopagebreak}
  {\par\nobreak\vfil\penalty0\vfilneg
   \vtop\bgroup}
  {\par\xdef\tpd{\the\prevdepth}\egroup
   \prevdepth=\tpd}

\definecolor{darkgreen}{rgb}{0,0.4,0}
\newcommand{\captioncode}{\scriptsize\usefont{T1}{DejaVuSansMono-TLF}{m}{n}}

\begin{document}
\allowdisplaybreaks[0]

\begin{absolutelynopagebreak}
  \begin{frontmatter}
    \begin{flushright}
      IPPP/16/36 \\
      QFET-2016-07 \\
      SI-HEP-2016-13 \\
      TUM-HEP-1043/16
    \end{flushright}
    \title{Near-threshold production of heavy quarks with
      \texttt{QQbar\_threshold}}
    \author[Munich]{M.~Beneke}
    \author[KEK]{Y.~Kiyo}
    \author[Durham]{A.~Maier}
    \author[Siegen,Bern]{J.~Piclum}
    \address[Munich]{
      Physik Department T31,
      James-Franck-Stra\ss{}e,
      Technische Universit\"at M\"unchen,
      D-85748 Garching, Germany
    }
    \address[KEK]{
      Department of Physics,
      Juntendo University,
      Inzai,
      Chiba 270-1695, Japan
    }
    \address[Durham]{
      Institute~for~Particle~Physics~Phenomenology,
      Department~of~Physics,
      Durham~University,
      Durham~DH1~3LE,
      United~Kingdom
    }
    \address[Siegen]{
      Theoretische Physik 1,
      Naturwissenschaftlich-Technische Fakult\"at,
      Universit\"at Siegen,
      57068 Siegen,
      Germany
    }
    \address[Bern]{
      Albert Einstein Center for Fundamental Physics,
      Institute for Theoretical Physics,
      University of Bern,
      3012 Bern,
      Switzerland
    }
    \begin{abstract}
      We describe the \texttt{QQbar\_threshold} library for computing the
      production cross section of heavy quark-antiquark pairs near
      threshold at electron-positron colliders. The prediction includes
all presently known QCD, electroweak, Higgs, and nonresonant corrections
in the combined nonrelativistic and weak-coupling expansion.
    \end{abstract}
    \begin{keyword}
      Perturbative calculations, Quantum Chromodynamics, Heavy Quarks
      \PACS 12.38.Bx, 14.65.-q
    \end{keyword}
  \end{frontmatter}
\end{absolutelynopagebreak}

\newpage

\section*{Program summary}
\begin{itemize}
\item \textit{Program title:} \texttt{QQbar\_threshold}
\item \textit{Programming language:} C++, Wolfram Language.
\item \textit{Computer:} PC.
\item \textit{Operating System:} Linux, OS X.
\item \textit{License:} GNU GPLv3.
\item \textit{RAM:} 60 MB
\item \textit{Disk Space:} 63 MB (26 MB download)
\item \textit{External Routines:} Boost~(\url{http:///www.boost.org}),\\
  GSL~(\url{http://www.gnu.org/software/gsl/}).
\item \textit{Nature of problem:}
  Precision predictions for the pair-production cross section
  near threshold are essential in order to determine the properties of heavy
  quarks.
\item \textit{Solution method:}
  Formulas for all known perturbative corrections are implemented, so
  that \texttt{QQbar\_threshold} provides a state-of-the-art theory
  prediction.
\item \textit{Restrictions:} Non-perturbative effects are not accounted
  for. This limits the applicability in the case of bottom quarks and excludes all
  lighter quarks. Due to the nonrelativistic approximation predictions for
  the cross section are only reliable near threshold.
\item \textit{Running time:} Typically about $5\,$ms per parameter point.
\end{itemize}

\newpage

\tableofcontents

\section{Introduction}
\label{sec:intro}

One of the main physics goals of envisaged high-energy electron-positron
colliders is to precisely measure the properties of the top quark. It is
expected that the top quark mass and width can be determined with high
accuracy by measuring the shape of the top-antitop production cross
section around
threshold~\cite{Seidel:2013sqa,Simon:2016htt}. Due to strong
non-perturbative effects such an analysis is not possible for the
lighter quarks observed at present low-energy electron-positron
colliders. For bottom quarks, however, $\Upsilon$ sum rules can be used
to extract the mass from moments of the pair-production cross section
near threshold~\cite{Novikov:1976tn,Novikov:1977dq,Voloshin:1987rp}. In both cases, a
precise theory prediction of the cross section is indispensable.

Near the production threshold, the Coulomb interaction between the quark
and the antiquark leads to a strong enhancement of the cross section,
and has to be included to all orders in perturbation theory. This is
achieved in the effective theory frameworks of potential nonrelativistic
quantum chromodynamics (PNRQCD)~\cite{Pineda:1997bj} and velocity nonrelativistic
quantum chromodynamics~\cite{Luke:1999kz}.  Corrections from
strong interactions up to next-to-next-to leading order (N${}^2$LO) have been
known for more than a decade~\cite{Hoang:2000yr} and are available in
both formalisms. More recently, also the calculation of the third-order
QCD corrections within PNRQCD has been
finished~\cite{Beneke:2015kwa}. Furthermore, corrections from P-wave
production~\cite{Beneke:2013kia}, non-resonant
production~\cite{Beneke:2010mp,Penin:2011gg,Jantzen:2013gpa,Ruiz-Femenia:2014ava},
Higgs
effects~\cite{Strassler:1990nw,Guth:1991ab,Harlander:1995dp,Eiras:2006xm,Beneke:2015lwa},
and further electroweak
interactions~\cite{Grzadkowski:1986pm,Guth:1991ab,Hoang:2004tg,Hoang:2006pd}
are known. While all of these parts are available, it is non-trivial to
combine all formulas consistently and evaluate the result numerically.

The \texttt{QQbar\_threshold} library provides functions to compute the
production cross section of heavy quark pairs near threshold and related
quantities like S-wave binding energies and bound state residues. It is
intended to be as flexible as possible, supporting a plethora of options
and tunable input parameters. All of the functionality documented in
this work can be accessed easily from both C++ and Wolfram Mathematica
programs. In the following we give an overview of the library and
its main functionality. An up-to-date comprehensive documentation can be
found on
\url{https://qqbarthreshold.hepforge.org/}. After
a short description of the installation process in
section~\ref{sec:install}, we explain the basic usage with some examples
in section~\ref{sec:examples}. Section~\ref{sec:xs_struct} describes the
structure of the cross section, which allows us to give a more detailed
account of all optional settings in section~\ref{sec:options}. We then
proceed to discuss some more advanced applications in section~\ref{sec:advanced_usage}. Finally,
section~\ref{sec:grid_gen} describes the generation of auxiliary grids.

\section{Installation}
\label{sec:install}

\subsection{Linux}
\label{sec:install-linux}

The easiest way to install \texttt{QQbar\_threshold} is via the included
installation script. The following software has to be available on the
system:
\begin{itemize}
\item a compiler complying with the C++11 standard (e.g. g++ 4.8 or
higher);
\item cmake (\href{http://cmake.org/}{\tt http://cmake.org/});
\item GSL, the GNU scientific library
(\href{http://www.gnu.org/software/gsl}{\tt
http://www.gnu.org/software/gsl});
\item the QCD library (included in the installation script);
\item the QCD library requires odeint from the boost libraries
(\href{http://www.boost.org/}{\tt http://www.boost.org/}).
\end{itemize}
It is recommended to run the installation script in a separate build
directory, e.g. \texttt{/tmp/build/}. After changing to such a
directory, the following code can be run in a terminal to download
\texttt{QQbar\_threshold} and install it to the directory \texttt{/my/path/}:
\begin{lstlisting}[language=sh]
wget https://www.hepforge.org/archive/qqbarthreshold/install.tar.gz
tar xzf install.tar.gz
cd QQbar_threshold_source
./install.sh /my/path/
\end{lstlisting}
If \texttt{/my/path/} is omitted, a default directory (usually
\texttt{/usr/local/}) will be used.

During installation there is the opportunity to change some
predefined physical constants like the W and Z masses and the default
settings for some of the options discussed in
section~\ref{sec:options}. A table of the default values is given in
appendix~\ref{sec:const}. Changing these values \emph{after installation} will
have no effect at best and might even lead to inconsistent results. If
Mathematica is available on the system\footnote{More precisely, the
\texttt{math} program to start the Mathematica command line interface
must be in the executable path.} the \texttt{QQbarThreshold} package
will also be installed automatically.

Before using the C++ library part, it may also be necessary to adjust
certain environment variables. After installation to the base directory
\texttt{/my/path/} the following settings are recommended:
%\begin{verbatim}
\begin{lstlisting}[language=sh]
LIBRARY_PATH="/my/path/lib:$LIBRARY_PATH"
LD_LIBRARY_PATH="/my/path/lib:$LD_LIBRARY_PATH"
CPLUS_INCLUDE_PATH="/my/path/include:$CPLUS_INCLUDE_PATH"
\end{lstlisting}
%\end{verbatim}

\subsection{OS X}
\label{sec:install-mac}

Under OS X, \texttt{QQbar\_threshold} can be installed like under Linux,
apart from two exceptions. First, the environment variable
\texttt{DYLD\_LIBRARY\_PATH} should be set in place of
\texttt{LD\_LIBRARY\_PATH}. Second, typical Mathematica installations do
not provide the required \texttt{math} executable and the
\texttt{QQbarThreshold} Mathematica package will not be installed
automatically. One way around this is to locate the
\texttt{WolframKernel} (or \texttt{MathKernel}) executable included in the
Mathematica installation and provide a small wrapper script. Assuming
\texttt{WolframKernel} can be found under
\texttt{/Applications/\allowbreak Mathematica.app/\allowbreak
  Contents/\allowbreak MacOS/} the following
shell script can be used:
\begin{lstlisting}[language=sh]
#!/bin/sh

MATH_PATH=/Applications/Mathematica.app/Contents/MacOS/
DYLD_LIBRARY_PATH="$DYLD_LIBRARY_PATH:$MATH_PATH"

$MATH_PATH/WolframKernel "$@"

\end{lstlisting}
For the installation of the Mathematica package to work, the script file
has to be in the executable path and must be named \texttt{math}. After
installing the \texttt{QQbarThreshold} Mathematica package, the
above script is no longer required and can be safely removed.

\section{Basic usage and examples}
\label{sec:examples}

In this section, we give a brief overview over
\texttt{QQbar\_threshold}'s main functionality and show several
code examples. For the sake of a more accessible presentation we
postpone the discussion of most details to later sections.

The main observables that can be computed with \texttt{QQbar\_threshold}
are the total cross section and the energy levels and residues of
quarkonium bound states. The cross section is calculated in picobarn,
whereas all other dimensionful quantities are given in (powers of)
GeV. While the examples below demonstrate the usage of specialised C++
header files, we also provide a header {QQbar\_threshold.hpp} which
exposes all functionality offered by the \texttt{QQbar\_threshold}
library. Functions related to $t\bar{t}$ production start with the
prefix \texttt{ttbar\_}; correspondingly \texttt{bbbar\_} designates
$b\bar{b}$ functions. All (public) parts of the library are in the
\texttt{QQbar\_threshold} namespace.

The C++ examples below have to be compiled with a reasonably recent
compiler (complying with the C++11 standard) and linked to the
\texttt{QQbar\_threshold} library. For example, one
could compile the first code snippet \texttt{resonance.cpp} with the g++
compiler (version 4.8 or higher) with the command
\begin{lstlisting}[language=sh]
g++ -o resonance -std=c++11 resonance.cpp -lQQbar_threshold
\end{lstlisting}
and run it with
\begin{lstlisting}[language=sh]
./resonance
\end{lstlisting}
The code for all examples will also be installed alongside with the
library. Assuming again \texttt{/my/path} as the base directory, it
can be found under
\texttt{/my/\allowbreak path/\allowbreak include/\allowbreak QQbar\_threshold/\allowbreak examples}.

\subsection{Mathematica usage}
\label{sec:math}

While the following main text generally describes the usage in C++
programs, the C++ code examples are also followed by equivalent
Mathematica code. After loading the package with
\lstinline[language=Mathematica]!Needs["QQbarThreshold`"]! an overview
over the available symbols can be obtained with
\lstinline[language=Mathematica]!Names["QQbarThreshold`*"]!. \texttt{QQbarThreshold}
follows the usual conventions for documenting symbols; e.g.
\lstinline[language=Mathematica]!Information[TTbarXSection]! or
\lstinline[language=Mathematica]!?TTbarXSection! explains the
\lstinline[language=Mathematica]!TTbarXSection! function and
\lstinline[language=Mathematica]!Options[TTbarXSection]!
shows its options and their default settings.

\subsection{Resonances}
\label{sec:resonances}

\texttt{QQbar\_threshold} can calculate the binding energy of the S-wave
$Q\bar{Q}$ resonances.  As a first example, we calculate the binding
energy of the $\Upsilon(1S)$ resonance at leading order. The binding
energy $E^{\text{RS}}_1$ depends on the chosen renormalisation scheme
and is defined by $E^{\text{RS}}_1 = M_{\Upsilon(1S)} -
2\*m^{\text{RS}}_b$, where $m^{\text{RS}}_b$ is the bottom-quark mass in
the given scheme RS, and $M_{\Upsilon(1S)}$ refers to the theoretically
computed mass of the $\Upsilon(1S)$. By default, the quark masses are defined
in the PS-shift scheme (cf. section~\ref{sec:mass_schemes}). All masses
and energies are in GeV, so the output implies that at leading order
$E^{\text{PS}}_1 = 0.273409\,$GeV.
\begin{lstlisting}[language=C++]
#include <iostream>
#include "QQbar_threshold/energy_levels.hpp"

int main(){
  namespace QQt = QQbar_threshold;
  const double mb_PS = 4.5; // bottom mass in PS scheme
  double E_1 = QQt::bbbar_energy_level(
      1,          // principal quantum number
      mb_PS,      // renormalisation scale
      mb_PS,      // quark mass
      QQt::LO     // perturbative order
  );
  std::cout << E_1 << '\n';
}
\end{lstlisting}
\begin{lstlisting}[language=Mathematica]
Needs["QQbarThreshold`"];

With[
   {mbPS = 4.5, scale = 4.5},
   E1 = BBbarEnergyLevel[1, scale, mbPS, "LO"]
];
Print[E1];
\end{lstlisting}
The corresponding residue (see eq.~(\ref{eq:ZN_def}) for a definition)
can be computed in very much the same way using the
\lstinline[language=C++]!bbbar_residue! function in C++ or
\lstinline[language=Mathematica]!BBbarResidue! in Mathematica. The
calculation of the (scheme-dependent) nonrelativistic wave function at the origin is
slightly more involved and are covered in
section~\ref{sec:wave_function}.

Higher-order corrections can be included by changing the last argument
to \lstinline[language=C++]!QQt::NLO!,
\lstinline[language=C++]!QQt::N2LO!, or
\lstinline[language=C++]!QQt::N3LO!
(\lstinline[language=Mathematica]!"NLO"!,
\lstinline[language=Mathematica]!"N2LO"!, or
\lstinline[language=Mathematica]!"N3LO"! in Mathematica). It should be
noted that at the highest order, i.e.~N${}^3$LO, energy levels and residues
are only implemented for the six lowest resonances. At lower orders,
however, the principal quantum number can be an arbitrary positive
integer.

\subsection{Cross sections}
\label{sec:xs}

One of the most interesting phenomenological applications is the threshold
scan of the $t\bar{t}$ production cross section. The functions
\lstinline[language=C++]!ttbar_xsection! and
\lstinline[language=C++]!bbbar_xsection! compute the
cross sections
\begin{align}
  \label{eq:rt_def}
 \sigma_t(s) ={}& \sigma(e^+ e^- \to W^+W^-b\bar{b})\,,\\
  \label{eq:rb_def}
  \sigma_b(s) ={}& \sigma(e^+ e^- \to b\bar{b})\,,
\end{align}
in picobarn. The name
\lstinline[language=C++]!ttbar_xsection! is justified by the fact that
near the $t\bar{t}$ threshold $\sigma_t$ is dominated by the production and
subsequent decay of a $t\bar{t}$ pair near mass shell; see
section~\ref{sec:nr_xs} for details.  Here is an example for computing the
cross section at a single point at next-to-leading order for
$m_t^{\text{PS}}(20\,\text{GeV}) = 168\,$GeV and a top width of $1.4\,$GeV with the
result $\sigma_t\big((340\,\text{GeV})^2\big) = 0.724149\,\text{pb}$:
\begin{lstlisting}[language=C++]
#include <iostream>
#include "QQbar_threshold/xsection.hpp"

int main(){
  namespace QQt = QQbar_threshold;
  std::cout
    // QQt::ttbar_xsection(sqrt_s, {scales}, {mass, width}, order)
    << QQt::ttbar_xsection(340., {80., 350.}, {168., 1.4}, QQt::NLO)
    << '\n';
}
\end{lstlisting}
\begin{lstlisting}[language=Mathematica]
Needs["QQbarThreshold`"];

Print[TTbarXSection[340., {80., 350.}, {168., 1.4}, "NLO"]];
\end{lstlisting}
Note that in this example two scales appear. As before, the first scale
is the overall renormalisation scale. The second scale is due to the
separation of resonant and nonresonant contributions to the cross
section (cf. section~\ref{sec:nr_xs}). It exclusively appears
in the top cross section, all other functions only take a single scale
as their argument.

\subsubsection{Grids}
\label{sec:grid}

For N${}^2$LO and N${}^3$LO corrections to the cross section, it is
necessary to load a precomputed grid first. Some default grids can be
found in the \texttt{grids} subdirectory of the installation. The
generation of custom grids is explained in
section~\ref{sec:grid_gen}. For convenience, the
\lstinline[language=C++]!grid_directory! function returns the directory
containing the default grids. The following code performs a threshold
scan at N${}^3$LO and prints a table of the cross sections for
centre-of-mass energies between $330$ and $345\,$GeV:
\begin{lstlisting}[language=C++]
#include "QQbar_threshold/load_grid.hpp"
#include "QQbar_threshold/xsection.hpp"
#include <iostream>

int main(){
  namespace QQt = QQbar_threshold;
  QQt::load_grid(QQt::grid_directory() + "ttbar_grid.tsv");
  const double mu = 50.;
  const double mu_width = 350.;
  const double mt_PS = 168.;
  const double width = 1.4;
  for (double sqrt_s = 330.; sqrt_s < 345.; sqrt_s += 1.0) {
    std::cout << sqrt_s << '\t'
       << QQt::ttbar_xsection(
           sqrt_s ,{mu, mu_width} ,{mt_PS , width}, QQt::N3LO
       )
       << '\n';
  }
}
\end{lstlisting}
\begin{lstlisting}[language=Mathematica]
Needs["QQbarThreshold`"];

LoadGrid[GridDirectory <> "ttbar_grid.tsv"];
With[
   {
     mu = 50.,
     mtPS = 168.,
     width = 1.4,
     muWidth = 350.,
     order = "N3LO"
   },
   Do[
      Print[
         sqrts, "\t",
         TTbarXSection[sqrts, {mu, muWidth}, {mtPS, width}, order]
      ],
      {sqrts, 330., 345., 1.}
   ]
];
\end{lstlisting}
Grids cover a specific range in the rescaled energy and width
coordinates
\begin{align}
  \label{eq:grid_coordinates}
  \tilde{E} ={}& - 4/(\alpha_s^2\*C_F^2\*m_Q)\*E\,,\\
  \tilde{\Gamma} ={}& - 4/(\alpha_s^2\*C_F^2\*m_Q)\*\Gamma\,,
\end{align}
where $m_Q$ is the pole quark mass, $E = \sqrt{s} - 2\*m_Q$ the kinetic
energy, and $C_F = 4/3$. If the arguments of the cross section functions
lead to rescaled coordinates outside the range covered by the grid an
exception
is thrown. In the Mathematica package an error message is
displayed and the cross section function will return a symbolic
\lstinline[language=Mathematica]!LibraryFunctionError!.

After loading a grid, its coordinate range can be identified as shown in
the following example. For the default top grid, the range is
$-10.4956 \le \tilde{E} \le 17.0554$ and $-1.83673 \le \tilde{\Gamma}
\le -0.918367$. For the default bottom grid we would find $-37.1901 \le
\tilde{E} \le -9.29752 \times 10^{-6}$ and $-9.29752 \times 10^{-8} \le
\tilde{\Gamma} \le -4.55432 \times 10^{-15}$
\begin{lstlisting}[language=C++]
#include "QQbar_threshold/load_grid.hpp"
#include "QQbar_threshold/xsection.hpp"
#include <iostream>

int main(){
  namespace QQt = QQbar_threshold;
  QQt::load_grid(QQt::grid_directory() + "ttbar_grid.tsv");
  auto range = QQt::grid_range();
  std::cout << "Et range: (" << range.Et_min
            << ", " << range.Et_max << ")\n";
  std::cout << "Gammat range: (" << range.Gammat_min
            << ", " << range.Gammat_max << ")\n";
}
\end{lstlisting}
\begin{lstlisting}[language=Mathematica]
Needs["QQbarThreshold`"];

LoadGrid[GridDirectory <> "ttbar_grid.tsv"];
Print["Et range: ", {GridEtMin[], GridEtMax[]}];
Print["Gammat range: ", {GridGammatMin[], GridGammatMax[]}];
\end{lstlisting}
If no grid is loaded $0$ will be returned for all coordinate limits.

Note that at most one grid can be used at any given time; if a second
grid is loaded it will replace the first one. Loading a grid is not
thread safe, i.e. one should not try to load more than one grid in
parallel. See section~\ref{sec:parallel} for more details on
parallelisation.

\subsubsection{Thresholds}
\label{sec:thresholds}

For convenience, there are also the functions
\lstinline[language=C++]!ttbar_threshold! and
\lstinline[language=C++]!bbbar_threshold!to compute the na\"ive
production threshold, which is given by twice the pole mass. In the
following example, a scan around this na\"ive threshold is performed.
The centre-of-mass energy ranges from $2\*m_t - 3\,$GeV to $2\*m_t +
5\,$GeV, where the pole mass $m_t$ is calculated from the input mass in
the PS scheme.
\begin{lstlisting}[language=C++]
#include <iostream>
#include "QQbar_threshold/load_grid.hpp"
#include "QQbar_threshold/xsection.hpp"

int main(){
  namespace QQt = QQbar_threshold;
  QQt::load_grid(QQt::grid_directory() + "ttbar_grid.tsv");
  const double mu = 50.;
  const double mu_width = 350.;
  const double mt_PS = 168.;
  const double width = 1.4;
  const auto order = QQt::N3LO;
  const double thr = QQt::ttbar_threshold(mu, mt_PS, order);
  for(double E = -3.; E < 5.; E += 1.0){
    double sqrt_s = thr + E;
    std::cout << sqrt_s << '\t'
      << QQt::ttbar_xsection(
          sqrt_s, {mu, mu_width}, {mt_PS, width}, order
      )
      << '\n';
  }
}
\end{lstlisting}
\begin{lstlisting}[language=Mathematica]
Needs["QQbarThreshold`"];

LoadGrid[GridDirectory <> "ttbar_grid.tsv"];
With[
   {
     mu = 50.,
     muWidth = 350.,
     mtPS = 168.,
     width = 1.4,
     order = "N3LO"
   },
   With[
      {thr = TTbarThreshold[mu, mtPS, order]},
      Do[
         Print[
            thr + energy, "\t",
            TTbarXSection[
               thr + energy, {mu, muWidth}, {mtPS, width}, order
            ]
         ],
         {energy, -3., 5., 1.}
      ]
   ]
];
\end{lstlisting}
For the case that the centre-of-mass energy is not required anywhere
else in the program, the C++ \lstinline[language=C++]!threshold! class
allows a more concise notation. If used inside the first argument of a
cross section function, it will be equivalent to a matching
\lstinline[language=C++]!ttbar_threshold! or
\lstinline[language=C++]!bbbar_threshold! function with the arguments of
the surrounding cross section function. Thus, in this example the
$b\bar{b}$ cross section at threshold $\sigma_b(4\*m_b^2) = 167.214\,\text{pb}$ is
evaluated:
\begin{lstlisting}[language=C++]
#include <iostream>
#include "QQbar_threshold/load_grid.hpp"
#include "QQbar_threshold/xsection.hpp"
#include "QQbar_threshold/threshold.hpp"

int main(){
  namespace QQt = QQbar_threshold;
  QQt::load_grid(QQt::grid_directory() + "bbbar_grid.tsv");
  std::cout
    << bbbar_xsection(QQt::threshold()+1e-3, 4.5, 4.5, QQt::N3LO)
    // equivalent to
    // << QQt::bbbar_xsection(
    //     QQt::bbbar_threshold(4.5, 4.5, QQt::N3LO) + 1e-3,
    //     4.5, 4.5, QQt::N3LO
    // )
    << '\n';
}
\end{lstlisting}
In Mathematica, we can use the symbol
\lstinline[language=Mathematica]!QQbarThreshold! to the same effect.
\begin{lstlisting}[language=Mathematica]
Needs["QQbarThreshold`"];

LoadGrid[GridDirectory <> "bbbar_grid.tsv"];
Print[BBbarXSection[QQbarThreshold + 10^-3, 4.5, 4.5, "N3LO"]];
\end{lstlisting}
Note that for the bottom quark the width is considered to be zero and
thus the continuum cross section is discontinuous at the production
threshold. In fact, the expression for the cross section directly at
threshold is currently not known to N$^3$LO and it is necessary to add a
small positive offset to the energy, which was somewhat arbitrarily set
to $1\,$MeV in the above examples.

\subsection{Basic options}
\label{sec:basic_options}

The functions provided by \texttt{QQbar\_threshold} support a plethora
of optional settings to control their behaviour. In this section we only
discuss a small selection; a short summary of all options
(table~\ref{tab:C++_to_math}) followed by a comprehensive discussion is
given in section~\ref{sec:options}. The default settings for
the options are given by \lstinline[language=C++]!top_options()! for
top-related functions and \lstinline[language=C++]!bottom_options()! for
the bottom-related counterparts.

As an example let us have another look at the $t\bar{t}$ threshold
scan. Here, we change the renormalisation scheme for the mass from the
default PS-shift scheme to the \MS{} scheme and discard
all Standard Model corrections beyond QCD.
\begin{lstlisting}[language=C++]
#include <iostream>
#include "QQbar_threshold/load_grid.hpp"
#include "QQbar_threshold/xsection.hpp"

int main(){
  namespace QQt = QQbar_threshold;
  QQt::load_grid(QQt::grid_directory() + "ttbar_grid.tsv");
  const double mu = 50.;
  const double mu_width = 350.;
  const double mt_MS = 160.; // mass mt(mt) in the MSbar scheme
  const double width = 1.4;
  QQt::options opt = QQt::top_options();
  // MSbar scheme with mu_MSbar = mt_MS
  opt.mass_scheme = {QQt::MSshift, mt_MS};
  // turn off all non-QCD corrections
  opt.beyond_QCD = QQt::SM::none;
  for(double sqrt_s = 330.; sqrt_s < 345.; sqrt_s += 1.0){
    std::cout << sqrt_s << '\t'
      << QQt::ttbar_xsection(
          sqrt_s, {mu, mu_width}, {mt_MS, width}, QQt::N3LO,
          opt
      )
      << '\n';
  }
}
\end{lstlisting}
\begin{lstlisting}[language=Mathematica]
Needs["QQbarThreshold`"];

LoadGrid[GridDirectory <> "ttbar_grid.tsv"];
With[
   {
     mu = 50.,
     muWidth = 350.,
     mtMS = 160., (* mass mt(mt) in the MSbar scheme *)
     width = 1.4,
     order = "N3LO"
   },
   Do[
      Print[
         sqrts, "\t",
         TTbarXSection[
            sqrts, {mu, muWidth}, {mtMS, width}, order,
            MassScheme -> {"MSshift", mtMS},
            BeyondQCD -> None
         ]
      ],
      {sqrts, 330., 345., 1.}
   ]
];
\end{lstlisting}
Another useful option allows to modify the reference
value for the strong coupling at the scale of the Z boson mass:
\begin{lstlisting}[language=C++]
#include <iostream>
#include "QQbar_threshold/load_grid.hpp"
#include "QQbar_threshold/xsection.hpp"

int main(){
  namespace QQt = QQbar_threshold;
  QQt::load_grid(QQt::grid_directory() + "ttbar_grid.tsv");
  const double mu = 50.;
  const double mu_width = 350.;
  const double mt_PS = 168.;
  const double width = 1.4;
  QQt::options opt = QQt::top_options();
  opt.alpha_s_mZ = 0.1174;
  for( double sqrt_s = 330.; sqrt_s < 345.; sqrt_s += 1.0){
    std::cout << sqrt_s << '\t'
       << QQt::ttbar_xsection(
           sqrt_s, {mu, mu_width}, {mt_PS, width}, QQt::N3LO,
           opt
       )
       << '\n';
  }
}
\end{lstlisting}
\begin{lstlisting}[language=Mathematica]
Needs["QQbarThreshold`"];

LoadGrid[GridDirectory <> "ttbar_grid.tsv"];
With[
   {
     mu = 50.,
     muWidth = 350.,
     mtPS = 168.,
     width = 1.4,
     order = "N3LO"
   },
   Do[
      Print[
         sqrts, "\t",
         TTbarXSection[
            sqrts, {mu, muWidth}, {mtPS, width}, order,
            alphaSmZ -> 0.1174
         ]
      ],
      {sqrts, 330., 345., 1.}
   ]
];
\end{lstlisting}
For debugging purposes it can be quite helpful to print the current
option settings. The following example shows the default options for
top-related functions. Note that for some options the default is
signified by a physically meaningless sentinel value.
\begin{lstlisting}[language=C++]
#include <iostream>
#include "QQbar_threshold/parameters.hpp"

int main(){
  std::cout << QQbar_threshold::top_options();
}
\end{lstlisting}
In the Mathematica package the current option settings if different from
the default are always visible through their specification in the
function call. The default settings for a given function can be inspected
with \lstinline[language=Mathematica]!Options[function]!.

It is also possible to inspect the settings used internally for the
actual calculation:
\begin{lstlisting}[language=C++]
#include <iostream>
#include "QQbar_threshold/parameters.hpp"

int main(){
  namespace QQt = QQbar_threshold;
  const double sqrt_s = 340.;
  const double mu = 50.;
  const double mu_width = 350.;
  const double mt_PS = 168.;
  const double width = 1.4;
  const auto order = QQt::N3LO;
  // internal settings for top cross section
  std::cout << QQt::top_internal_settings(
      sqrt_s, {mu, mu_width}, {mt_PS, width}, order,
      QQt::top_options()
  );
  // internal settings for energy levels and residues
  std::cout << QQt::top_internal_settings(
      mu, mt_PS, order,
      QQt::top_options()
  );
}
\end{lstlisting}
\begin{lstlisting}[language=Mathematica]
Needs["QQbarThreshold`"];

With[
   {
      sqrts = 340.,
      mu = 50.,
      muWidth = 350.,
      mtPS = 168.,
      width = 1.4,
      order = "N3LO"
   },
   Print[
      TopInternalSettings[sqrts, {mu, muWidth}, {mtPS, width}, order]
   ];
   Print[TopInternalSettings[mu, mtPS, order]];
];
\end{lstlisting}
Correspondingly, for bottom quarks the function
\lstinline[language=C++]!bottom_internal_settings! can be used. The
arguments match the respective energy level, residue, or cross section
function.

\subsection{Scheme conversion}
\label{sec:scheme_conv}

While the PS scheme is appropriate for the description of threshold
observables, in other kinematic regions schemes like \MS{} may be more
suitable. For converting masses to the pole scheme,
\texttt{QQbar\_threshold} provides the functions
\lstinline[language=C++]!top_pole_mass! and
\lstinline[language=C++]!bottom_pole_mass!. As shown in the following
example, using these functions iteratively then allows conversions
between arbitrary schemes. For the input PS mass $m_t^\text{PS}(20\,\text{GeV}) =
168\,$GeV we find a pole mass of $m_t = 169.827\,$GeV and an \MS{} mass
of $m_t^{\MS}(m_t^{\MS}) = 160.035\,$GeV.
\begin{lstlisting}[language=C++]
#include <iostream>
#include "QQbar_threshold/scheme_conversion.hpp"

int main(){
  namespace QQt = QQbar_threshold;
  const double mt_PS = 168.;
  const double mu = 50.;
  //convert to pole scheme
  const double mt_Pole = QQt::top_pole_mass(mu, mt_PS, QQt::N3LO);
  //convert to MSbar scheme
  const double precision = 1e-4;
  QQt::options opt = QQt::top_options();
  double mt_MS = mt_PS;
  double delta_M;
  do{
    opt.mass_scheme = {QQt::MSshift, mt_MS};
    delta_M = QQt::top_pole_mass(mu, mt_MS, QQt::N3LO, opt) - mt_Pole;
    mt_MS -= delta_M;
  } while(std::abs(delta_M) > precision);
  std::cout << "pole mass: " << mt_Pole << '\n'
            << "MSbar mass: " << mt_MS << '\n';
}
\end{lstlisting}
\begin{lstlisting}[language=Mathematica]
Needs["QQbarThreshold`"];

With[
   {mtPS = 168., mu = 50., precision = 10^-4},
   (* convert to pole scheme  *)
   mtPole = TopPoleMass[mu, mtPS, "N3LO"];
   (* convert to MSbar scheme *)
   mtMS = mtPS;
   For[
      deltaM = TopPoleMass[
         mu, mtMS, "N3LO", MassScheme -> {"MSshift", mtMS}
      ] - mtPole,
      Abs[deltaM] > precision,
      deltaM = TopPoleMass[
         mu, mtMS, "N3LO", MassScheme -> {"MSshift", mtMS}
      ] - mtPole,
      mtMS -= deltaM
   ];
   Print["pole mass: ", mtPole];
   Print["MSbar mass: ", mtMS];
];
\end{lstlisting}

\subsection{Top quark width}
\label{sec:top_width}

The width of the top quark is an external parameter of the
\lstinline[language=C++]!ttbar_xsection! function and independent of
other input parameters like the mass and the strong coupling. In order
to ensure consistency with the Standard Model prediction, the
\lstinline[language=C++]!top_width! function can be used. It is
recommended to always use the highest available order (i.e. N${}^2$LO)
for the top width, even if the cross section is computed at a lower
order. The following code computes a width of $1.36003\,$GeV from the
given input:
\begin{lstlisting}[language=C++]
#include <iostream>
#include "QQbar_threshold/width.hpp"

int main(){
  namespace QQt = QQbar_threshold;
  const double mt_PS = 171.5;
  const double mu = 50.;
  std::cout << QQt::top_width(mu, mt_PS, QQt::N2LO) << '\n';
}
\end{lstlisting}
\begin{lstlisting}[language=Mathematica]
Needs["QQbarThreshold`"];

With[
   {
      mtPS = 171.5,
      mu = 50.
   },
   Print[TopWidth[mu, mtPS, "N2LO"]];
];
\end{lstlisting}
It should be noted that \lstinline[language=C++]!top_width! behaves
very differently from the other functions contained in
\texttt{QQbar\_threshold}.  Since N${}^3$LO corrections are unknown the
order is limited to N${}^2$LO. However, internally the input mass is
converted to the pole mass using N${}^3$LO conversion regardless of the
order argument of the \lstinline[language=C++]!top_width! function. At
N${}^2$LO, an approximation of the electroweak corrections based on the
exact results of~\cite{Denner:1990ns,Eilam:1991iz} is included in
addition to the QCD corrections~\cite{Jezabek:1993wk, Fischer:2000kx,
Blokland:2005vq}. More precisely, the electroweak corrections are
assumed to be a flat $1.7\%$ of the leading-order width. In contrast to
all other functions, we use the Fermi constant $G_F$ instead of the
running QED coupling as input parameter. Finally, a number of options available
for other functions are ignored. In particular, the electroweak
corrections cannot be turned off or altered in any way and the bottom
quark is always assumed to be massless.

\section{Structure of the cross section}
\label{sec:xs_struct}

Before discussing the optional settings in detail, we first give an
overview over the structure of the cross section as defined in
eqs.~(\ref{eq:rt_def}), (\ref{eq:rb_def}) up to N${}^3$LO in PNRQCD. A
more detailed account of the effective field theory framework is given
in~\cite{Beneke:2013jia,Beneke:2015}.

\subsection{Power counting}
\label{sec:poco}

In PNRQCD, an expansion in $\alpha_s \sim v \ll 1$ is performed, where
$v = [\sqrt{s}/m_Q - 2]^{1/2}$ is the non-relativistic velocity of the
quarks and $m_Q$ their pole mass. The Coulomb interaction leads to terms
scaling with powers of $\alpha_s/v \sim 1$, which are resummed to all
orders. Concerning the electroweak interactions including the Higgs
boson, we choose the power counting $\alpha \sim y_t^2 \sim \alpha_s^2$
for the QED coupling constant and the top Yukawa coupling. We include
pure QCD corrections up to N${}^3$LO, i.e. order $\alpha_s^3 \sim \alpha_s^2
v \sim \alpha_s v^2 \sim v^3$ relative to the leading-order cross
section. Similarly, Higgs corrections are considered to the same
order $\alpha_s y_t^2$,  and the Higgs mass counts as a hard scale, i.e.~$m_H\sim m_t$. The remaining electroweak corrections are mostly
only included at lower orders, as detailed in the following.

\subsection{Resummation of QED effects}
\label{sec:QED_resum}

It is customary to absorb large logarithmic corrections due to vacuum
polarisation into a running QED coupling constant $\alpha(\mu_\alpha)$,
which coincides with the fine structure constant $\alpha \equiv
\alpha(0)$ in the Thomson limit. The total cross section $\sigma(e^+
e^- \to q\bar{q})$ is then proportional to $\alpha(\mu_\alpha)^2$. We
therefore factorise the cross sections $\sigma_Q$ with $Q=b,t$ defined
in eqs.~(\ref{eq:rt_def}), (\ref{eq:rb_def}) as follows.
\begin{equation}
  \label{eq:sigma_noQED}
  \sigma_Q = \frac{4\*\pi\*\alpha(\mu_\alpha)^2}{3\*s}\*R_Q\,.
\end{equation}
$R_Q$ then depends on the QED coupling only through higher-order corrections.

A further source of large logarithms is given by photon initial state
radiation off the electron--positron pair. Currently, we exclude this
correction and all other QED corrections to the initial state.

\subsection{Nonresonant cross section}
\label{sec:nr_xs}

Since for top quarks the width is non-negligible, it is necessary to
consider the full process $e^+ e^- \to W^+ W^- b\bar{b}$ instead of just
the production of an on-shell $t\bar{t}$ pair. A systematic analysis in
the framework of unstable particle effective
theory~\cite{Beneke:2003xh,Beneke:2004km} shows that the cross section
can then be written as the sum of resonant and non-resonant production:
\begin{equation}
  \label{eq:sigma_nonres}
  R_Q(s) = R_{\text{res}}(s) + R_{\text{non-res}}(s)\,.
\end{equation}
While the resonant part by construction only contains the contributions
from top quarks near their mass shell, the invariant mass of the final
state $W\,b$ pair in the nonresonant part can be quite
different from the top quark mass. In order to reduce such background
contributions, it is possible to specify a cut on the invariant mass
(see section~\ref{sec:options}). The current implementation in
\texttt{QQbar\_threshold} only includes the NLO~\cite{Beneke:2010mp}
nonresonant cross section.

Both the resonant and the non-resonant part are separately divergent. We
remove the poles using $\MS$ subtraction and associate the remaining
logarithms with a new scale $\mu_w$ (cf. section~\ref{sec:xs}). While
these logarithms cancel order by order in the sum
(eq.~(\ref{eq:sigma_nonres})), a dependence on $\mu_w$ remains in the
present implementation at N$^2$LO and N$^3$LO, since the N$^2$LO and
N$^3$LO corrections to the non-resonant cross section are still
unknown. However, it has already been checked~\cite{Jantzen:2013gpa}
that the logarithms indeed cancel at N$^2$LO once the N$^2$LO
non-resonant contribution is included.

During the evaluation of the nonresonant cross section, interpolation on
a precomputed grid is performed. While physical values of the W and
the top quark mass are covered by a built-in grid, exotic parameter
settings may require the generation of a custom nonresonant grid. This
is covered in section~\ref{sec:grid_gen}. Custom grids can be loaded
with \lstinline[language=C++]!load_nonresonant_grid(gridfile)! (or, in
Mathematica, \lstinline[language=Mathematica]!LoadNonresonantGrid[gridfile]!).

\subsection{Production channels}
\label{S_P_wave}

While the resonant quark pair is mostly produced in an S wave, there is
also a subleading P-wave contribution starting at N${}^2$LO. Thus, the
resonant cross section can be decomposed as
\begin{equation}
  \label{eq:sigma_S_P}
  R_{\text{res}}(s) = R_S(s) + R_P(s)\,.
\end{equation}
$R_S$ and $R_P$ can be expressed in terms of the imaginary parts of the vector and
axialvector polarisation functions, respectively. One obtains
\begin{align}
  \label{eq:sigma_S}
  R_S(s) ={}& R_{S, \text{QCD}}(s) +
  R_{S, \text{EW}}(s)\,,\displaybreak[0]\\
  \label{eq:sigma_S_QCD}
  R_{S, \text{QCD}}(s) ={}& \big[{C^{(v)}}^2 +
  {C^{(a)}}^2\big]\*12\*\pi\*\im[\Pi_{\text{PR}}^{(v)}(s)]
  \,,\displaybreak[0]\\
  \label{eq:sigma_P}
  R_P(s) ={}& a_Q^2\*\big[a_e^2 +
  v_e^2\big]\*\frac{s^2}{(s-m_Z^2)^2}\*12\*\pi\*\im[\Pi_{\text{PR}}^{(a)}(s)]\,,\displaybreak[0]\\
  \label{eq:Cp_v}
  C^{(v)} ={}& e_e\*\*e_Q + v_Q\,\*v_e\*\frac{s}{s - m_Z^2}\,,\displaybreak[0]\\
  \label{eq:Cp_a}
  C^{(a)} ={}& -v_Q\,\*a_e\*\frac{s}{s - m_Z^2}\,,
\end{align}
where $v_f$ is the vector coupling of a fermion to the Z boson and $a_f$
the corresponding axialvector coupling given by
\begin{equation}
  \label{eq:v_f_a_f}
  v_f = \frac{T_3^f - 2\*e_f\*s_w^2}{2\*s_w\*c_w}\,, \qquad a_f = \frac{T_3^f}{2\*s_w\*c_w}\,.
\end{equation}
 $e_f$ is the fermion charge in units of the positron charge, $T_3^f$
its third isospin component, $c_w = m_W/m_Z$ the
cosine of the Weinberg angle, and $s_w = (1 -
c_w^2)^{1/2}$. $R_{S, \text{EW}}(s)$ is the electroweak
correction to S-wave
production~\cite{Grzadkowski:1986pm,Guth:1991ab,Hoang:2004tg,Hoang:2006pd};
more details are given in section~\ref{sec:sigma_EW}.

\subsection{Pole resummation}
\label{sec:pole_resum}

The polarisation functions exhibit poles at $E = E_N - i\Gamma$, where
$E = \sqrt{s} - 2m_Q$ is the kinetic energy, $\Gamma$ the quark width,
and $E_N$ the (real) binding energy of the $N$th bound state. For
reasons detailed in~\cite{Beneke:1999qg} the pole contributions to the
polarisation functions should be resummed by subtracting the
contribution expanded around the leading-order pole position and adding
back the unexpanded contributions, i.e.
\begin{equation}
  \label{eq:pole_resum}
  \Pi_{\text{PR}}^{(v)}(s) = \Pi^{(v)}(s) + \frac{N_C}{2\*m_Q^2} \sum_{N=1}^\infty\bigg\{
  \frac{\big[Z_N\big]_{\text{expanded}}}{\big[E_N - E - i\Gamma\big]_{\text{unexpanded}}} -
  \bigg[\frac{Z_N}{E_N - E - i\Gamma}\bigg]_{\text{expanded}}\bigg\}\,,
\end{equation}
and similar for the axialvector polarisation function
$\Pi_{\text{PR}}^{(a)}(s)$ with the P-wave energy levels $E_N^P$ and residues $Z_N^P$.
A more precise definition of $Z_N$ is given in
section~\ref{sec:EN_ZN}. It should be emphasised that in the limit of a
vanishing width, i.e. for bottom quarks, pole resummation has no effect
on the (continuum) cross section.

In the actual implementation, it is of course
not possible to evaluate the sum in eq.~(\ref{eq:pole_resum}) up to
infinity. The number of resummed poles is instead set via an option of
the cross section functions and defaults to 6. From the scaling
of the residues with $N$ the resulting error on the cross section can be
estimated to be comparable to the difference between resumming 4 and 6
poles and is typically at most about 2 per mille.

\subsection{Hard matching}
\label{sec:matching}

\subsubsection{QCD and Higgs}
\label{sec:QCD_Higgs_matching}

The polarisation functions without pole resummation are a product of
hard current matching coefficients and the non-relativistic Green
functions $G, G^P$,
\begin{align}
  \label{eq:Pi_v}
  \Pi^{(v)}(s) ={}& \frac{2\*N_c}{s}\*c_v\*\bigg[c_v -
  \frac{E + i \Gamma}{m_Q}\frac{d_v}{3}\bigg]\*G(E) + \dots\,, \displaybreak[0]\\
  \label{eq:Pi_a}
\Pi^{(a)}(s) ={}& \frac{2\*N_c}{m_Q^2\*s}\frac{d-2}{d-1}\*c_a^2\*G^P(E) + \dots\,.
\end{align}
Here $G(E)$ denotes the Green function at complex energy $E + i\*\Gamma$.
It is also understood that the products are consistently expanded to
N${}^3$LO and higher-order terms are dropped. The perturbative
expansions of the matching coefficients of the non-relativistic currents
up to the required order can be put into the following form:
\begin{align}
  \label{eq:hard_matching_exp}
  c_v ={}& 1 + \frac{\alpha_s(\mu)}{4\pi} c_v^{(1)}
+ \bigg(\frac{\alpha_s(\mu)}{4\pi}\bigg)^2 c_v^{(2)}
+ \bigg(\frac{\alpha_s(\mu)}{4\pi}\bigg)^3 c_v^{(3)}\notag\\
&+ \frac{y_Q^2}{2}\bigg[c_{vH}^{(2)} + \frac{\alpha_s(\mu)}{4\pi}\*c_{vH}^{(3)}\bigg]\,,\\
  \label{eq:dv_exp}
d_v ={}& d_v^{(0)} + \frac{\alpha_s(\mu)}{4\pi} d_v^{(1)}\,,\\
  \label{eq:ca_exp}
c_a ={}& 1 + \frac{\alpha_s(\mu)}{4\pi} c_a^{(1)}\,,
\end{align}
where numerically $d_v^{(0)} = 1$. The index $H$ indicates corrections
where a Higgs boson couples exclusively to the heavy
quark. $\alpha_s(\mu)$ denotes the strong coupling constant in the \MS{}
scheme at the overall renormalisation scale $\mu$. Explicit formulas for
the coefficients can be found in~\cite{Eiras:2006xm,Beneke:2013jia,Marquard:2014pea}.

\subsubsection{Electroweak}
\label{sec:sigma_EW}

The electroweak correction in eq.~(\ref{eq:sigma_S}) can be written (up
to the NNLO order considered here) as
\begin{equation}
  \label{eq:sigma_EW} R_{S, \text{EW}}(s) =
\frac{12\*N_c}{s}\*\alpha(\mu_\alpha)\*c_v\*\im\Big[(C^{(v)}\*C_{\text{EW}}^{(v)}
+ C^{(a)}\*C_{\text{EW}}^{(a)})\,\*G_{\text{PR}}(E)\Big] + \dots\,.
\end{equation}
In contrast to the QCD and Higgs hard matching coefficients discussed in
section~\ref{sec:QCD_Higgs_matching} the electroweak Wilson coefficients
$C_{\text{EW}}^{(v)}$ and $C_{\text{EW}}^{(a)}$ have a non-vanishing
imaginary part that contributes to the cross section. They can be
decomposed further into a pure QED contribution and corrections
involving at least one $W$, $Z$, or Goldstone boson\footnote{We also
  include the Higgs-loop correction to the s-channel $Z$ propagator in
  $C^{(v, a)}_{\text{WZ}}$.
}
\begin{equation}
  \label{eq:Cva_EW}
  C^{(v, a)}_{\text{EW}} =
  C^{(v, a)}_{\text{QED}} + C^{(v, a)}_{\text{WZ}}\,.
\end{equation}
Note that $C_{\text{WZ}}^{(v)}$ and $C_{\text{WZ}}^{(a)}$ do not contain
corrections from Higgs bosons coupling exclusively to heavy quarks;
these are instead absorbed into $c_v$
(cf. eq.~(\ref{eq:hard_matching_exp})). As already mentioned in
section~\ref{sec:QED_resum}, $C_{\text{QED}}^{(v)}$ and
$C_{\text{QED}}^{(a)}$ do not include purely photonic corrections that
couple only to the initial-state leptons, yet.

 $G_{\text{PR}}(E)$ is the pole-resummed Green function (see also
section~\ref{sec:pole_resum}) given by
\begin{equation}
  \label{eq:PR_GF}
  G_{\text{PR}}(E) = G(E) + \sum_{N=1}^\infty\bigg\{
  \frac{\big[|\psi_N(0)|^2\big]_{\text{expanded}}}{\big[E_N - E - i\Gamma\big]_{\text{unexpanded}}} -
  \bigg[\frac{|\psi_N(0)|^2}{E_N - E - i\Gamma}\bigg]_{\text{expanded}}\bigg\}\,,
\end{equation}
where $\psi_N(0)$ is the quarkonium wave function at the origin. Like in
the QCD pole resummation (eq.~(\ref{eq:pole_resum})), $E_N$ denotes the
binding energy to the same order as the total cross section,
i.e. N${}^2$LO or N${}^3$LO.

According to the power counting outlined in section~\ref{sec:poco}, the
electroweak correction first contributes at N${}^2$LO. N${}^3$LO
contributions arise from QCD corrections to either of $c_v$, $G(E)$, or
$C^{(v, a)}_{\text{EW}}$. Since the corrections to
$C^{(v, a)}_{\text{EW}}$ are not known completely, we
currently only include the N${}^3$LO contributions due to corrections to
$c_v$ and $G(E)$.

Since the mass of the bottom quark lies significantly below the
electroweak scale, corrections due to W, Z, and Higgs bosons should be
considered in an effective (Fermi) theory, if at all. For this reason we
discard all corrections contributing to
$C^{(v, a)}_{\text{WZ}}$ when computing the bottom
production cross section. Technically, these corrections are excluded
whenever $m_Q < m_Z$, so for unphysical values of the top mass also the
top production cross section would be affected.

\subsection{Green functions}
\label{sec:GF}

The expansion of the S-wave Green function to third order can be written as
\begin{equation}
  \begin{split}
    \label{eq:GF_exp}
    G(E) ={}& \langle \mathbf{0} | \hat{G}_0(E) |\mathbf{0}\rangle +
    \langle \mathbf{0} | \hat{G}_0(E)\,i\,\delta V\, i\,\hat{G}_0(E)
    |\mathbf{0}\rangle
    \\
    &+ \langle \mathbf{0} | \hat{G}_0(E)\,i\,\delta V\,
    i\,\hat{G}_0(E)\,i\,\delta V\, i\,\hat{G}_0(E) |\mathbf{0}\rangle\\
    &+ \langle \mathbf{0} | \hat{G}_0(E)\,i\,\delta V\,
    i\,\hat{G}_0(E)\,i\,\delta V\, i\,\hat{G}_0(E)\,i\,\delta V\,
    i\,\hat{G}_0(E) |\mathbf{0}\rangle\\
    &+\delta^{us}G(E) + \dots\,.
  \end{split}
\end{equation}
Here, $\hat{G}_0(E)$ is the Green function operator of unperturbed
PNRQCD and $\delta V$ denotes a correction to the leading-order
potential. $\delta^{us}G(E)$ stands for the ultrasoft correction
contributing only at third order. Again, eq.~(\ref{eq:GF_exp}) is to be
understood as consistent expansions to N${}^3$LO.

For the P-wave Green function an analogous formula holds. In this case,
there is an additional insertion of $p \cdot p'$, where $p$ and $p'$ are
the momenta of the initial and final state. Only the first two terms (no
perturbation and a single insertion) are needed at N${}^3$LO,
cf.~\cite{Beneke:2013kia} for details.

\subsection{Potentials}
\label{sec:potentials}

The corrections to the potential up to third order can be classified in
the following way:
\begin{equation}
  \label{eq:delta_V}
  \delta V = \underbrace{\delta_C V + \delta_{\text{QED}} V}_{\text{NLO}} + \underbrace{\delta_{1/r^2} V + \delta_{\delta} V + \delta_p
  V  + \delta \text{kin}}_{\text{N}^2\text{LO}} + \underbrace{\delta_H V}_{\text{N}^3\text{LO}}\,,
\end{equation}
\begin{itemize}
\item $\delta_C V$: Corrections to the colour Coulomb potential.
\item $\delta_{\text{QED}} V$: QED Coulomb potential.
\item $\delta_{1/r^2} V$: Potential proportional to $1/r^2$
  (equivalently $1/m$).
\item $\delta_{\delta} V$: Potential proportional to $\delta(r)$
  (equivalently $1/m^2$).
\item $\delta_{p} V$: Momentum-dependent potential.
\item $\delta_H V$: Potential due to Higgs exchange.
\item $\delta \text{kin}$: Kinetic energy correction.
\end{itemize}
The braces indicate the order at which these potential corrections first
appear. While all QCD corrections to the potentials are implemented up
to N${}^3$LO, we generally do not include N${}^3$LO electroweak
corrections for the sake of consistency with the electroweak corrections
to the hard matching discussed in section~\ref{sec:sigma_EW}.

\begin{figure}
  \centering
  \begin{tabular}{ccccccc}
    \includegraphics{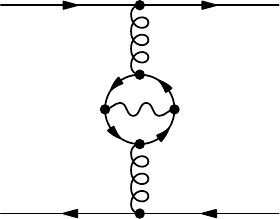} && \includegraphics{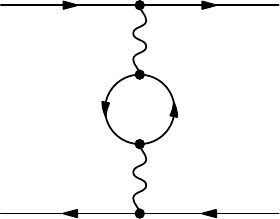}
    && \includegraphics{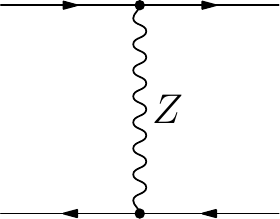} && \includegraphics{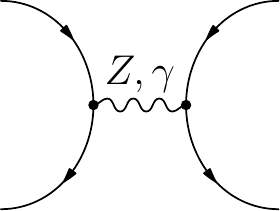}\\
a) && b) && c) && d)
  \end{tabular}
  \caption{Third-order potential corrections that are not included in
    \texttt{QQbar\_threshold}. From left to right: a) QED correction to
    the colour Coulomb potential; b) one-loop correction to the QED
    Coulomb potential; c) $Z$ potential; d) annihilation followed by
    pair production.}
  \label{fig:pot_missing}
\end{figure}

In this vein, we exclude the QED corrections to both the colour Coulomb
potential (cf. figure~\ref{fig:pot_missing}) and the delta potential
$\delta_{\delta} V$. For $\delta_{\text{QED}} V$,
the one-loop (order $\alpha^2$) QED contribution, corresponding to a
N${}^3$LO correction, is also excluded. Together with other
N${}^3$LO electroweak corrections we neglect the potential induced by
the exchange of $Z$ bosons. Note that $W$ exchange is formally beyond
third order according to our power counting. Finally, the
nonrelativistic quark pair can annihilate into a virtual photon or $Z$
that again produces a nonrelativistic quark pair. This also constitutes
a N${}^3$LO electroweak correction and is therefore not taken into
account. Note that we do include multiple insertions of the QED Coulomb
potential $\delta_{\text{QED}} V$ into the Green function (see
eq.~(\ref{eq:GF_exp})). For the similar case of the colour Coulomb
potential, this prescription has been shown to lead to better agreement
with numerical solutions to the Schr\"odinger
equation~\cite{Beneke:2005hg}.

For the case of a non-zero light quark (e.g. charm) mass, the potentials
receive further contributions, which are known to
N${}^2$LO~\cite{Melles:1998dj,Melles:2000dq,Hoang:2000fm}. Up to this
order, only the colour Coulomb potential is affected. It can be decomposed as
\begin{equation}
  \label{eq:delta_ml}
  \delta_C V = \delta_{C,0} V + \delta_{C,m_l} V\,,
\end{equation}
where $\delta_{C,0} V$ corresponds to the contribution for a vanishing
light quark mass $m_l$.

Because of the strong mass hierarchy, corrections to top-related
observables due to a non-zero light-quark mass are negligible. However,
in the case of the bottom quark the charm-quark mass is of the same
order as the heavy-quark momentum. According to our power counting
(section~\ref{sec:poco}) the charm contributions to the colour Coulomb
potential are therefore formally of the same order as the contributions
from massless quarks.

In practice, charm-mass effects are found to be numerically small but
computationally rather expensive, typically requiring numerical
Mellin-Barnes integrations. Therefore, we liberally discard sub-leading
effects during calculations. For the continuum cross section, at most
the single insertion of the potential $\delta_{C, m_l} V$ at NLO into
the S-wave Green function is taken into account (cf. first line of
eq.~(\ref{eq:GF_exp})). The bound state energies and residues also
contain the single insertion of the N${}^2$LO potential and the double
insertion of the NLO potential~\cite{Beneke:2014pta}.\footnote{This
  double insertion correction is available, but not included by
default. See the option \lstinline[basicstyle=\captioncode,language=C++]!double_light_insertion!
in section~\ref{sec:options}.} Note that for the
pole resummation eqs.~(\ref{eq:pole_resum}), (\ref{eq:PR_GF}) also at
most the single insertion of $\delta_{C, m_l} V$ at NLO is considered.

\subsection{Energy levels and residues}
\label{sec:EN_ZN}

As noted in section~\ref{sec:pole_resum}, the S-wave energy levels $E_N$
are given by the position of the poles in the vector polarisation
function, or, equivalently, the S-wave Green function $G(E)$. The
residues of the Green function then correspond to the modulus squared of
the wave function at the origin:
\begin{equation}
  \label{eq:psi_N_def}
  G(E) \xrightarrow{E+i\Gamma \to E_N} \frac{|\psi_N(0)|^2}{E_N - E - i \Gamma}\,.
\end{equation}
Since $\psi_N(0)$ is a factorisation scheme dependent quantity, the
\lstinline[language=C++]!bbbar_residue! and
\lstinline[language=C++]!ttbar_residue! functions instead compute $Z_N$,
which is defined as
\begin{equation}
  \label{eq:ZN_def}
  Z_N = \frac{4\*m_Q^2}{s_N}\*c_v\*\bigg[c_v -
  \frac{E_N}{m_Q}\frac{d_v}{3}\bigg]\*|\psi_N(0)|^2
\end{equation}
with $s_N = (2\*m_Q + E_N)^2$.

\subsection{Mass schemes}
\label{sec:mass_schemes}

So far, all formulas have been expressed in terms of the pole mass of
the heavy quark. Mass values in other schemes RS can be converted
via relations of the form
\begin{equation}
  \label{eq:conv_pole}
  m_Q = m_Q^{\text{RS}} + \sum_{i=0}^{o} \delta m_i^{\text{RS}}\,,
\end{equation}
where $0 \leq o \leq 3$ is the considered order according to the PNRQCD
power counting summarised in section~\ref{sec:poco}. The pole mass can
now be substituted in two different ways~\cite{Beneke:2015}:
\begin{itemize}
\item First compute the numerical value of the pole mass from the mass
  value in the scheme RS by using relation~(\ref{eq:conv_pole}). The value
  obtained for the pole mass in this way will strongly depend on the order
  $o$ used in equation~(\ref{eq:conv_pole}). Then evaluate the expressions
  for the cross section, residues, and energy levels in the \emph{pole}
  scheme with the pole mass $m_Q$ as determined above. This the {\it
    shift} prescription.
\item Symbolically replace the pole mass in all expressions by the mass
  in the scheme RS via relation~(\ref{eq:conv_pole}). Then perform a
  systematic expansion wherever $\delta m_i^{\text{RS}}$ constitutes a
  small correction. To ensure both a consistent expansion and
  order-by-order renormalon cancellation, $\delta m_0^{\text{RS}}$ has to
  be of order $E\sim v^2$. Finally, insert the numeric value of the mass
  in the scheme RS. This corresponds to the {\it insertion}
  prescription.

  Deviating from this general rule, in the present version of the code the
  residue $Z_N$ in the insertion scheme is computed as $(m_Q^{\text RS})^2
  \times [Z_N/m_Q^2]_{\text{RS}}$ where the quantity in square brackets is
  transformed from the pole scheme to RS according to the general
  rule. That is, we perform the naive replacement $m_Q \to
  m_Q^{\text{RS}}$ \emph{without} the correction terms $\delta
  m_i^{\text{RS}}$ in the factors $4\*m_Q^2/s_N$ in eq.~(\ref{eq:ZN_def})
  and also in $N_C/(2\*m_Q^2)$ in eq.~(\ref{eq:pole_resum}).  This has no
  effect on the cross section, since $m_Q$ cancels in the product of these
  factors. However, $Z_N$ as defined above differs slightly from the value
  $[Z_N]_{\text{RS}}$ it would attain, if the general rule were applied to
  $Z_N$ directly.

  The insertion prescription leads to unphysical oscillations of the
  cross section near threshold~\cite{Beneke:2015}. What is more, for the
  bottom quark the cross section is only defined in the sense of a
  distribution. This is due to the expansion of the threshold step
  function $\theta(s - 4\*m_b^2)$ in $\delta m_i^{\text{RS}} < \sqrt{s}$
  for $i > 0$. It is therefore recommended to use the shift prescription
  instead.
\end{itemize}
In both prescriptions, the energy variable $E^\text{RS}$ is
defined as $E^\text{RS} = \sqrt{s} - 2\*m_Q^\text{RS}$, and similarly
the binding energies are defined by the bound state masses minus
$2\*m_Q^\text{RS}$ so that $s_N = (2\*m_Q^\text{RS} + E_N^\text{RS})^2$
is scheme independent (see also section~\ref{sec:resonances}).

So far, apart from the pole scheme, the following schemes are
implemented in \texttt{QQbar\_threshold}:
\begin{itemize}
\item The potential-subtracted (PS) scheme~\cite{Beneke:1998rk} up to
N${}^3$LO~\cite{Beneke:2005hg}. Corrections from a non-zero light-quark
mass are only contained up to N${}^2$LO~\cite{Beneke:2014pta}. We define
the subtraction potential to not include any electroweak
corrections. Because of this, the first-order QED corrections leads
to a visible shift of the $t\bar{t}$ cross section peak for fixed input
PS mass, but contrary to QCD, higher-order QED and electroweak
corrections are rapidly convergent.
\item The 1S scheme~\cite{Hoang:1998ng}. Up to N${}^3$LO the conversion formula
to the pole scheme (cf. eq.~(\ref{eq:conv_pole})) is given by
\begin{equation}
  \label{eq:1S_to_pole}
  m_Q = m_Q^{\text{1S}} - \frac{E_1(m_Q)}{2} = m_Q^{\text{1S}} -
  \frac{E_1(m_Q^{\text{1S}})}{2} + \frac{E_1(m_Q^{\text{1S}})}{4}
  \frac{\partial E_1(m_Q^{\text{1S}})}{\partial m_Q^{\text{1S}}} + \dots\,.
\end{equation}
 Since the 1S scheme is closely connected to the bound state energy
levels, corrections are implemented to the same order, i.e. N${}^3$LO
for the QCD and Higgs corrections and N${}^2$LO for the electroweak
corrections.
\item The $\MS$ scheme in QCD. For this scheme, we keep $\delta m_0^{\MS}$
at LO in eq.~(\ref{eq:conv_pole}), $\delta m_1^{\MS}$ at NLO, and so on. Corrections are
available at order $\alpha_s^4$~\cite{Marquard:2015qpa}, which
corresponds to N$^3$LO as required at the present highest
accuracy. Since $\delta m_0^{\MS}$ is of the same order as $v$ (rather
than $v^2$ as in the PS and 1S scheme), only the shift prescription is
self-consistent with the convention adopted here.  We also include
corrections from a non-zero light-quark mass to
N${}^2$LO~\cite{Bekavac:2007tk}. We define this scheme via the pure QCD
relation to the pole mass and therefore do not include any electroweak
corrections to the mass conversion.
\end{itemize}

\section{Options}
\label{sec:options}

Options are set by passing an \lstinline[language=C++]!options! struct
as the last function argument in the C++ library or by the conventional
\lstinline[language=Mathematica]!option -> value! arguments in the
Mathematica package. A short overview over all options is given in
table~\ref{tab:C++_to_math}. In the C++ case, it is recommended to
modify an object initialised with the helper functions
\lstinline[language=C++]!top_options()! or
\lstinline[language=C++]!bottom_options()! as demonstrated in
section~\ref{sec:basic_options}. It should be noted that exotic option
settings (especially for the \lstinline[language=C++]!contributions!
option) can easily lead to scheme-dependent or otherwise unphysical
results.

Note that in many cases there is more than one option that disables
certain parts. In case of conflicting settings, a contribution is
discarded. For example, if all QED contributions are switched off
through the \lstinline[language=C++]!beyond_QCD! option, setting
\mbox{\lstinline[language=C++]!contributions.v_QED_Coulomb[0] = 1.!} will
\emph{not} re-enable the QED Coulomb potential.

\begin{table}
  \centering

  \begin{tabularx}{\textwidth}{llX}
    \toprule
    C++ name                                         & Mathematica name                                       & Description                                                    \\
    \midrule
    \lstinline[language=C++]!contributions!          & \lstinline[language=Mathematica]!Contributions!        & Fine-grained control over higher-order corrections.            \\
    \lstinline[language=C++]!alpha_s_mZ!             & \lstinline[language=Mathematica]!alphaSmZ!             & Value of $\alpha_s(m_Z)$.                                      \\
    \lstinline[language=C++]!alpha_s_mu0!            & \lstinline[language=Mathematica]!alphaSmu0!            & Values of $\mu_0$ and $\alpha_s(\mu_0)$.                       \\
    \lstinline[language=C++]!m_Higgs!                & \lstinline[language=Mathematica]!mHiggs!               & Value of $m_H$.                                                \\
    \lstinline[language=C++]!Yukawa_factor!          & \lstinline[language=Mathematica]!YukawaFactor!         & Multiplier for heavy-quark Yukawa coupling.                    \\
    \lstinline[language=C++]!resonant_only!          & \lstinline[language=Mathematica]!ResonantOnly!         & Toggle for non-resonant contribution.                          \\
    \lstinline[language=C++]!invariant_mass_cut!     & \lstinline[language=Mathematica]!InvariantMassCut!     & Cut on $W\,b$ invariant mass.                                  \\
    \lstinline[language=C++]!ml!                     & \lstinline[language=Mathematica]!ml!                   & Value of light-quark mass.                                     \\
    \lstinline[language=C++]!r4!                     & \lstinline[language=Mathematica]!r4!                   & Value of parameter in N$^3$LO \MS{} to pole scheme conversion. \\
    \lstinline[language=C++]!alpha!                  & \lstinline[language=Mathematica]!alpha!                & Value of $\alpha(\mu_\alpha)$.                                 \\
    \lstinline[language=C++]!mu_alpha!               & \lstinline[language=Mathematica]!muAlpha!              & Value of scale $\mu_\alpha$ for QED coupling.                  \\
    \lstinline[language=C++]!resum_poles!            & \lstinline[language=Mathematica]!ResumPoles!           & Number of resummed poles.                                      \\
    \lstinline[language=C++]!beyond_QCD!             & \lstinline[language=Mathematica]!BeyondQCD!            & Toggle for higher-order corrections beyond QCD.                \\
    \lstinline[language=C++]!mass_scheme!            & \lstinline[language=Mathematica]!MassScheme!           & Mass renormalisation scheme.                                   \\
    \lstinline[language=C++]!production!             & \lstinline[language=Mathematica]!Production!           & Toggle for production channels.                                \\
    \lstinline[language=C++]!expand_s!               & \lstinline[language=Mathematica]!ExpandEnergyFactor!   & Toggle for expansion of $1/s$ prefactors.                      \\
    \lstinline[language=C++]!double_light_insertion! & \lstinline[language=Mathematica]!DoubleLightInsertion! & Toggle for double insertions of light-quark potential.         \\
    \bottomrule
  \end{tabularx}
  \caption{Members of the C++
\lstinline[basicstyle=\captioncode,language=C++]!options! structure and
equivalent Mathematica options.}
  \label{tab:C++_to_math}
\end{table}

The \lstinline[language=C++]!options! struct has the following members:
\begin{itemize}
\item \lstinline[language=C++]!contributions!: Specifies multiplicative
factors for the potentials (cf. eq.~(\ref{eq:delta_V}) and current
matching coefficients
(eqs.~(\ref{eq:hard_matching_exp}),~(\ref{eq:dv_exp})). For example,
setting
\begin{lstlisting}[language=C++]
options opt;
opt.contributions.v_delta = {{0., 1.}};
\end{lstlisting}
implies that corrections due to the leading-order delta potential are
discarded, but corrections from the next-to-leading delta potential are
kept, i.e. multiplied by $1$. Table~\ref{tab:potentials} lists the
relations to the definitions in eqs. (\ref{eq:delta_V}), (\ref{eq:hard_matching_exp}), (\ref{eq:dv_exp}).

In Mathematica, the \lstinline[language=Mathematica]!Contributions!
option expects a list of all contributions with their coefficients:
\begin{lstlisting}[language=Mathematica]
Contributions -> {
   vCoulomb -> {1., 1., 1.},
   vdelta -> {0., 1.},
   ...
}
\end{lstlisting}
To facilitate the usage, the \texttt{QQbarThreshold} package provides
the auxiliary functions
\lstinline[language=Mathematica]!ExceptContributions! and
\lstinline[language=Mathematica]!OnlyContributions! which set the
factors for all contributions that are not listed explicitly to $1$ or
$0$, respectively. For example, to discard only the leading-order delta
potential one could use
\begin{lstlisting}[language=Mathematica]
Contributions -> ExceptContributions[vdelta -> {0., 1.}]
\end{lstlisting}

\begin{table}
  \centering
  \begin{tabular}{llll}
    \toprule
    C++ name                                & Mathematica name                              & Corrections                                                       & Defined eq.          \\
    \midrule
    \lstinline[language=C++]!v_Coulomb!     & \lstinline[language=Mathematica]!vCoulomb!    & $\{\delta_CV^{(1)}{}^\dagger, \delta_CV^{(2)}, \delta_CV^{(3)}\}$ & (\ref{eq:delta_V})           \\
    \lstinline[language=C++]!v_delta!       & \lstinline[language=Mathematica]!vdelta!      & $\{\delta_\delta V^{(0)}, \delta_\delta V^{(1)}\}$                & (\ref{eq:delta_V})           \\
    \lstinline[language=C++]!v_r2inv!       & \lstinline[language=Mathematica]!vr2inv!      & $\{\delta_{1/r^2} V^{(1)}, \delta_{1/r^2} V^{(2)}\}$              & (\ref{eq:delta_V})           \\
    \lstinline[language=C++]!v_p2!          & \lstinline[language=Mathematica]!vp2!         & $\{\delta_{p} V^{(0)}, \delta_{p} V^{(1)}\}$                      & (\ref{eq:delta_V})           \\
    \lstinline[language=C++]!v_kinetic!     & \lstinline[language=Mathematica]!vkinetic!    & $\{\delta\text{kin}\}$                                            & (\ref{eq:delta_V})           \\
    \lstinline[language=C++]!ultrasoft!     & \lstinline[language=Mathematica]!ultrasoft!   & $\{\delta^{us}G(E)\}$                                             & (\ref{eq:GF_exp})            \\
    \lstinline[language=C++]!v_Higgs!       & \lstinline[language=Mathematica]!vHiggs!      & $\{\delta_H V^{(0)}\}$                                            & (\ref{eq:delta_V})           \\
    \lstinline[language=C++]!v_QED_Coulomb! & \lstinline[language=Mathematica]!vQEDCoulomb! & $\{\delta_{\text{QED}}V^{(0)}\}$                                  & (\ref{eq:delta_V})           \\
    \lstinline[language=C++]!cv!            & \lstinline[language=Mathematica]!cv!          & $\{c_v^{(1)}, c_v^{(2)}, c_v^{(3)}\}$                             & (\ref{eq:hard_matching_exp}) \\
    \lstinline[language=C++]!cv_Higgs!      & \lstinline[language=Mathematica]!cvHiggs!     & $\{c_{vH}^{(2)}, c_{vH}^{(3)}\}$                                  & (\ref{eq:hard_matching_exp}) \\
    \lstinline[language=C++]!Cv_QED!        & \lstinline[language=Mathematica]!CvQED!       & $\{C^{(v)}_{\text{QED}}\}$                                        & (\ref{eq:Cva_EW})            \\
    \lstinline[language=C++]!Ca_QED!        & \lstinline[language=Mathematica]!CaQED!       & $\{C^{(a)}_{\text{QED}}\}$                                        & (\ref{eq:Cva_EW})            \\
    \lstinline[language=C++]!Cv_WZ!         & \lstinline[language=Mathematica]!CvWZ!        & $\{C^{(v)}_{\text{WZ}}\}$                                         & (\ref{eq:Cva_EW})            \\
    \lstinline[language=C++]!Ca_WZ!         & \lstinline[language=Mathematica]!CaWZ!        & $\{C^{(a)}_{\text{WZ}}\}$                                         & (\ref{eq:Cva_EW})            \\
    \lstinline[language=C++]!dv!            & \lstinline[language=Mathematica]!dv!          & $\{d_v^{(0)},d_v^{(1)}\}$                                         & (\ref{eq:dv_exp})            \\
    \lstinline[language=C++]!ca!            & \lstinline[language=Mathematica]!ca!          & $\{c_a^{(1)}\}$                                                   & (\ref{eq:ca_exp})            \\
    \bottomrule
  \end{tabular}
  \caption{List of potential and matching coefficient corrections that
can be modified with the
\lstinline[basicstyle=\captioncode,language=C++]!contributions!
option. In general superscripts refer to the number of loops associated
with a correction. For $c_{vH}$ we instead follow the notation
of~\cite{Beneke:2015lwa}, where the superscript indicates the PNRQCD
order. ${}^\dagger\, \delta_CV^{(1)}$ multiplies both the contributions from
the NLO colour Coulomb potential and the QED Coulomb potential.}
  \label{tab:potentials}
\end{table}
\item \lstinline[language=C++]!alpha_s_mZ! or
  \lstinline[language=C++]!alpha_s_mu0! specifies the input value
for the strong coupling constant. If the option
\lstinline[language=C++]!alpha_s_mZ! is used, it is assumed that the
given value corresponds to $\alpha_s(m_Z)$.
\lstinline[language=C++]!alpha_s_mu0! specifies both a reference scale
and the value of $\alpha_s$ at that scale. For example
\begin{lstlisting}[language=C++]
options opt;
opt.alpha_s_mu0 = {10., 0.22};
\end{lstlisting}
sets $\alpha_s(10\,\text{GeV}) = 0.22$. If both options are set, the
value for \lstinline[language=C++]!alpha_s_mZ! is ignored.

The input value for the strong coupling is evolved automatically to
the overall renormalisation scale using four-loop evolution. For
bottom-related functions decoupling to the four-flavour theory is
performed only if the input scale is above the decoupling scale
\lstinline[language=C++]!mu_thr! defined in the \texttt{constants.hpp}
header. With the current default settings, decoupling is performed at
twice the scale-invariant mass $m_b^{\MS}(m_b^{\MS}) = 4.203\,$GeV. Note that for
top-related functions the input value for the strong coupling is always
assumed to refer to the five-flavour theory and no decoupling is
performed.

The final values used for the actual calculations can be inspected with
the \lstinline[language=C++]!alpha_s_bottom! and
\lstinline[language=C++]!alpha_s_top! functions from the header
alpha\_s.hpp (or \lstinline[language=C++]!alphaSBottom!,
\lstinline[language=C++]!alphaSTop! in Mathematica), which take the renormalisation scale as their first
argument and the value of
either \lstinline[language=C++]!alpha_s_mZ! or
\lstinline[language=C++]!alpha_s_mu0! as their second argument.
\item \lstinline[language=C++]!m_Higgs!: Specifies the value of the Higgs boson mass.
\item \lstinline[language=C++]!Yukawa_factor!: Specifies a multiplier
for the top-quark Yukawa coupling. This can be used to parametrise a
possible deviation from the Standard Model relation between the top-quark
mass and the coupling to the Higgs boson.

We assume that this deviation is caused by the dimension-6 operator
\begin{equation}
  \label{eq:yukawa_mod_op}
  \Delta {\cal L} = -
  \frac{c_{\text{NP}}}{\Lambda^2}(\phi^\dagger\phi)(\bar{Q}_3i\sigma^2\phi^* t_R) + \text{h.c.}\,,
\end{equation}
which implies the relation~\cite{Beneke:2015lwa}
\begin{equation}
\lstinline[language=C++]!Yukawa_factor! = 1+\frac{c_\text{NP}}{\Lambda^2}\frac{v^3}{\sqrt{2} m_t}\,.
\end{equation}
While this operator modifies the coupling to the physical Higgs boson,
the couplings to the Goldstone bosons remain unchanged, provided they
are expressed in terms of the top-quark mass. The operator also
generates four- and five-point vertices. Since we count the coupling
$c_\text{NP} v^2/\Lambda^2$ as N$^2$LO, similar to $\alpha$ and $y_t^2$
(c.f. section~\ref{sec:poco}), these vertices contribute only through a
Higgs tadpole diagram to the top self-energy, which has no effect in the
top mass renormalisation schemes adopted here. Hence in the present
approximation, the only effect of the dimension-6 operator is a
rescaling of the Yukawa coupling.
\item \lstinline[language=C++]!resonant_only!: If set to
  \lstinline[language=C++]!true!, the nonresonant contribution to the
  cross section (cf. eq.~(\ref{eq:sigma_nonres})) is discarded.
\item \lstinline[language=C++]!invariant_mass_cut!: Specifies an
invariant mass cut for the nonresonant contribution in
eq.~(\ref{eq:sigma_nonres}). The invariant mass of each $W\,b$ pair in
the final state is restricted to the region between $m_t -
\text{\lstinline[language=C++]!invariant_mass_cut!}$ and $m_t +
\text{\lstinline[language=C++]!invariant_mass_cut!}$. By default, the
loosest possible cut
$\text{\lstinline[language=C++]!invariant_mass_cut!} = m_t - m_W$ is
taken, which yields the total cross section.
\item \lstinline[language=C++]!ml!: Specifies the value of the light
(e.g. charm or bottom) quark mass. The input value should be the \MS{}
quark mass at the overall renormalisation scale $\mu$. This option only
affects the mass of light quarks in virtual corrections, the bottom
quarks in the final state of the process $e^+ e^- \to W^+ W^- b\bar{b}$
are always assumed to be massless.
\item \lstinline[language=C++]!r4!: Specifies the four-loop
coefficient for the conversion between the pole and the \MS{} scheme
(see eq.~(\ref{eq:conv_pole})). More precisely,
\lstinline[language=C++]!r4! is defined by the relation
\begin{equation}
  \label{eq:r4_def}
  m_Q = m_Q^{\MS}\big(m_Q^{\MS})\big)\*\big[1 + r_1\*a_s + r_2\*a_s^2 +
  r_3\*a_s^3 + \lstinline[language=C++]!r4!\,\*a_s^4 + {\cal O}\big(a_s^5\big)\big]\,,
\end{equation}
where $a_s = \alpha_s^{(n_l)}\big(m_Q^{\MS}\big)/\pi$. Note that only the
$n_l$ massless quark flavours contribute to the running of the strong
coupling; \lstinline[language=C++]!r4! therefore differs slightly from
the constant $c_m^{(4)}(\mu = m^{\MS}\big(m^{\MS}\big)$
of~\cite{Marquard:2015qpa}, which refers to the theory with $n_l + 1$
active flavours.

This option is only relevant if the \MS{} scheme was chosen and only
affects the conversion at N${}^3$LO.
\item \lstinline[language=C++]!alpha!: Specifies the value of the QED coupling
  constant at the scale \mbox{\lstinline[language=C++]!mu_alpha!.} This mainly
  affects the overall normalisation factor in eq.~(\ref{eq:sigma_noQED}),
  but also all electroweak corrections.
\item \lstinline[language=C++]!mu_alpha!: Specifies the scale for the QED coupling
  constant.
\end{itemize}
Let us comment on the usage of the two previous options by adopting two
  examples. (I) We want to compute the top cross section using a different
  value for $\alpha(m_Z)$, say $\alpha(m_Z) = 1/130$. (II) We think that the
  default scale $\mu_\alpha = m_Z$ is too low and want to use $\alpha(m_t)$ as
  input. In scenario (I) we set \lstinline[language=C++]!alpha = 1./130.! and we
  are done. In scenario (II), we first have to look up or compute
  $\alpha(m_t)$ elsewhere. For the sake of the argument, let us assume
  $\alpha(m_t) = 1/125$. Then we set \lstinline[language=C++]!alpha = 1./125.!
  and \lstinline[language=C++]!mu_alpha = !$m_t$. Setting
  \lstinline[language=C++]!mu_alpha! is necessary, because $\mu_\alpha$
  appears \emph{explicitly} (i.e. not only as an argument of $\alpha$) in
  the formula for the cross section.
\begin{itemize}
\item \lstinline[language=C++]!resum_poles!: Specifies the number of bound states that
  are resummed into the vector polarisation function and the electroweak
  contribution to the cross section
  (cf. eqs.~(\ref{eq:pole_resum}), (\ref{eq:PR_GF})). At N${}^3$LO the
  current maximum value is 6; at lower orders there is no such upper
  limit. This option does not affect the pole resummation for the
  axialvector polarisation function, where in the current version always
  the three leading poles are resummed.
\item \lstinline[language=C++]!beyond_QCD!: Specifies the Standard Model corrections
beyond QCD that should be taken into account. More precisely, each
setting defines the Lagrangian of the underlying full theory, from
which the higher-order corrections in the effective nonrelativistic
theory are then derived. For example, with the setting
\lstinline[language=C++]!beyond_QCD = SM::Higgs! corrections involving Higgs bosons
coupling to the heavy quarks are added to the usual PNRQCD
corrections. Note that this option does not affect the leading-order
production process, i.e. s-channel production via a virtual $Z$ or
photon is still taken into account with the above setting, although
higher-order corrections due to photons or $Z$ bosons are
disabled. Nonresonant production is also unaffected by this option.

The possible settings are shown in
figure~\ref{fig:beyond_QCD}. $\mathcal{L}_{\text{QCD}}$ and
$\mathcal{L}_{\text{SM}}$ denote the usual QCD and Standard Model
Lagrangians. Furthermore, we use
\begin{align}
  \label{eq:lagrangians}
  \mathcal{L}_{\text{QED}} ={}& -\frac{1}{4} F_{\mu\nu}F^{\mu\nu} +
  \sum_{l \in \text{leptons}} \overline{\psi}_l i \slashed{\partial}  \psi_l -
  \sum_{f \in \text{fermions}} e_f \overline{\psi}_f \slashed{A} \psi_f
   \,,\\
  \mathcal{L}_{\text{Higgs}} ={}&  \frac{1}{2} (\partial_\mu H)^2 -
  \frac{1}{2}\*m_H^2\*H^2 - \sqrt{\frac{\lambda}{2}}\*m_H\*H^3
- \frac{\lambda}{4}\*H^4
- \frac{y_t}{\sqrt{2}} \overline{t} t H\,,
\end{align}
where $\lambda = \frac{\pi\*\alpha\*m_H^2}{2\*m_W^2\*s_w^2}$.
\begin{figure}
  \centering
\includegraphics[width=\linewidth]{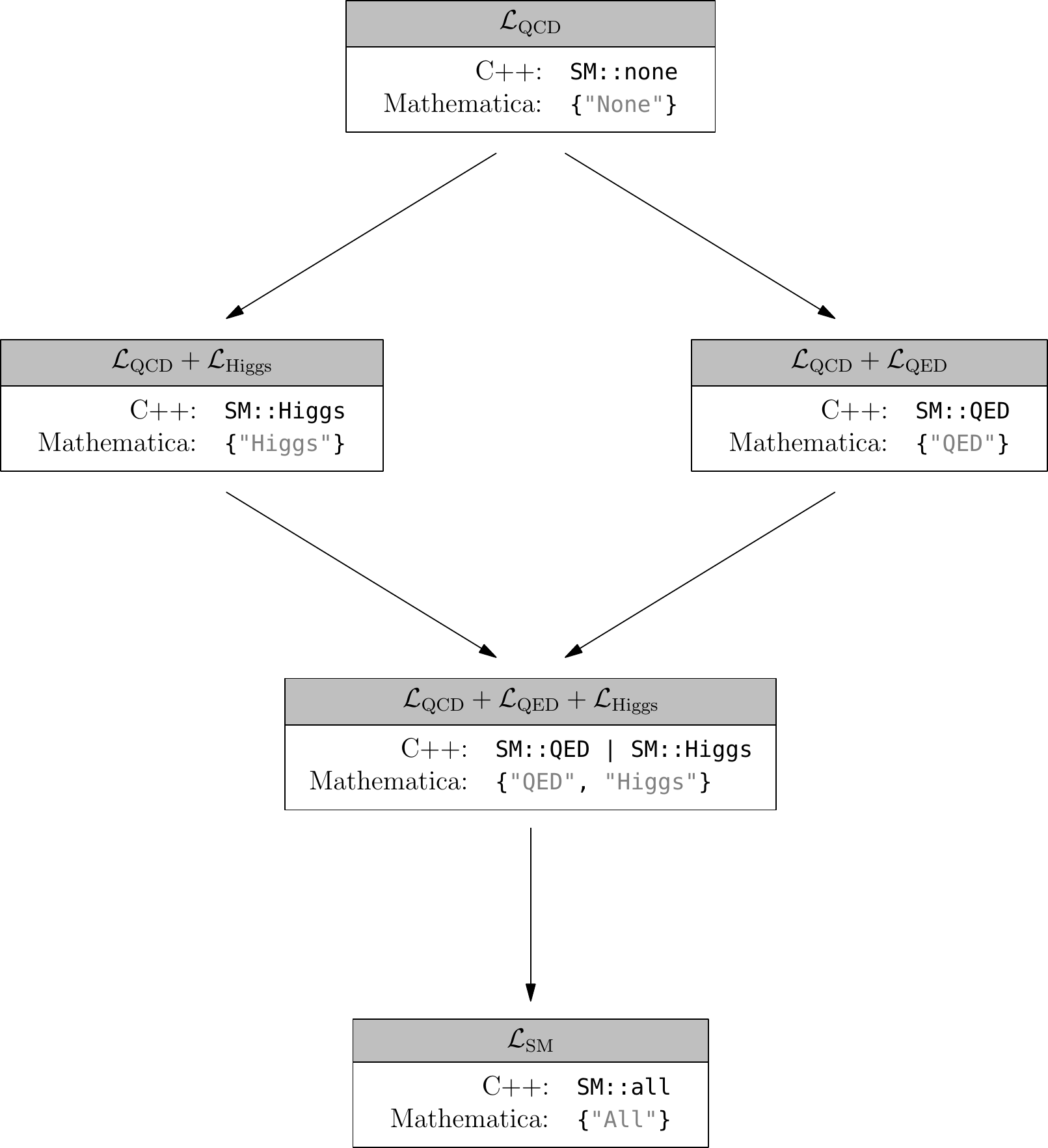}
  \caption{Possible settings for the
    \lstinline[basicstyle=\captioncode,language=C++]!beyond_QCD! option and the associated
    Lagrangians defining the perturbative corrections that are
    included. See eq.~(\ref{eq:lagrangians}) for a definition of
    $\mathcal{L}_{\text{QED}}$ and $\mathcal{L}_{\text{Higgs}}$.}
  \label{fig:beyond_QCD}
\end{figure}
\item \lstinline[language=C++]!mass_scheme!: Specifies the
renormalisation scheme and scale of the quark mass. Note that in the C++
library it is mandatory to specify a scale:
\begin{lstlisting}[language=C++]
options opt;
opt.mass_scheme = {PSshift, 20.};
opt.mass_scheme = {Pole, 0.};
\end{lstlisting}
For schemes without intrinsic scale (e.g. the pole scheme) the second
value can be set arbitrarily. In the Mathematica package, it can also be
omitted completely:
\begin{lstlisting}[language=Mathematica]
MassScheme -> {"PSshift", 20.}
MassScheme -> "Pole"
\end{lstlisting}
A list of the available schemes is given
in table~\ref{tab:schemes}.
\begin{table}
  \centering
  \begin{tabular}{lll}
    \toprule
    C++ name                            & Mathematica name                            & Description                               \\
    \midrule
    \lstinline[language=C++]!PS!        & \lstinline[language=Mathematica]!"PS"!      & The potential-subtracted insertion scheme \\
    \lstinline[language=C++]!PSshift!   & \lstinline[language=Mathematica]!"PSshift"! & The potential-subtracted shift scheme     \\
    \lstinline[language=C++]!OneS!      & \lstinline[language=Mathematica]!"1S"!      & The 1S insertion scheme                   \\
    \lstinline[language=C++]!OneSshift! & \lstinline[language=Mathematica]!"1Sshift"! & The 1S shift scheme                       \\
    \lstinline[language=C++]!Pole!      & \lstinline[language=Mathematica]!"Pole"!    & The pole scheme                           \\
    \lstinline[language=C++]!MSshift!   & \lstinline[language=Mathematica]!"MSshift"! & The \MS{} scheme                          \\
    \bottomrule
\end{tabular}
  \caption{List of available schemes. For details, see section~\ref{sec:mass_schemes}}
  \label{tab:schemes}
\end{table}
\item \lstinline[language=C++]!production!: Specifies which production
  channels are taken into account. The possible settings are:
  \begin{itemize}
  \item \lstinline[language=C++]!photon_only!: Production only via a
    virtual photon. This effectively discards the P-wave contribution from
    eq.~(\ref{eq:sigma_S_P}), the second term in the vector production
    operator, and the axialvector
    production operator defined in eqs.~(\ref{eq:Cp_v}) and
    (\ref{eq:Cp_a}). In addition, box corrections and corrections to the
    production via a virtual Z boson are discarded in the electroweak
    contribution to the cross section (eq.~(\ref{eq:sigma_EW})).
  \item \lstinline[language=C++]!S_wave_only!: Production only via an
    S-wave photon or Z, i.e. the P-wave contribution in
    eq.~(\ref{eq:sigma_S_P}) is discarded.
  \item \lstinline[language=C++]!all!: All possible production channels.
  \end{itemize}
  The corresponding Mathematica settings are
  \lstinline[language=Mathematica]!"PhotonOnly"!,
  \lstinline[language=Mathematica]!"SWaveOnly"!, and
  \lstinline[language=Mathematica]!"All"! (or \lstinline[language=Mathematica]!All!).
\item \lstinline[language=C++]!expand_s!: Specifies the treatment of the
overall factor $1/s$ in the polarisation functions defined in
eqs.~(\ref{eq:Pi_v}), (\ref{eq:Pi_a}) and the electroweak correction (eq.~(\ref{eq:sigma_EW})). If set to
\lstinline[language=C++]!true!, $s = (2\*m_Q + E)^2$ is expanded in
$E/m_Q \sim v^2 \ll 1$ to the appropriate order. This also affects the
prefactor $1/s_N$ in the residue $Z_N$ (eq.~(\ref{eq:ZN_def})).
\item \lstinline[language=C++]!double_light_insertion!: Specifies whether
double insertions (second line of eq.~(\ref{eq:GF_exp})) of the
light-quark potential correction $\delta_{C, m_l}V$ defined in
eq.~(\ref{eq:delta_ml}) are taken into account. This option only affects
the calculation of the energy levels and residues; in the continuum
cross section double insertions are always neglected.

The impact on observables that can be computed reliably within
perturbation theory is typically small. For example, in the
determination of the bottom quark mass from the 10th moment of the cross
section the end result is changed by about $0.1$ per mille, compared
to an overall change of around $0.5$ per mille when combined with the dominant
single insertions~\cite{Beneke:2014pta}. In infrared-sensitive
quantities like binding energies and residues of bound states with
principal quantum number $N > 5$ the effects can become significant.

The calculation of the double insertions is computationally very
expensive, requiring the evaluation of an infinite sum over integrals
(cf.~\cite{Beneke:2014pta}). Therefore, setting this option to
\lstinline[language=C++]!true! leads to a considerable slowdown by a
factor between $100$ and $1000$. In order to avoid an even more severe
slowdown as well as numerical instabilities, the current implementation
only computes the first few terms of the sum, so that the result is not
very precise. Furthermore, the implementation is not thread safe
(cf. section~\ref{sec:parallel}).
\end{itemize}
Finally, the default values for version 1.0 of \texttt{QQbar\_threshold}
are shown in table~\ref{tab:defaults}. Since default settings may change
in later versions, it is recommended to also consult the online
documentation under \url{https://qqbarthreshold.hepforge.org/}.

\begin{table}
  \centering
  \begin{tabular}{lll}
    \toprule
    Option                                           & Default for top                                                           & Default for bottom                                                           \\
    \midrule
    \lstinline[language=C++]!contributions!          & All set to 1                                                              & All set to 1                                                                 \\
    \lstinline[language=C++]!alpha_s_mZ!             & \lstinline[language=C++]!alpha_s_mZ = 0.1184!                             & \lstinline[language=C++]!alpha_s_mZ = 0.1184!                                \\
    \lstinline[language=C++]!alpha_s_mu0!            & (not set)                                                                 & (not set)                                                                    \\
    \lstinline[language=C++]!m_Higgs!                & \lstinline[language=C++]!m_Higgs = 125.!                                  & \lstinline[language=C++]!m_Higgs = 125.!                                     \\
    \lstinline[language=C++]!Yukawa_factor!          & \lstinline[language=C++]!1!                                               & \lstinline[language=C++]!0!                                                  \\
    \lstinline[language=C++]!resonant_only!          & \lstinline[language=C++]!false!                                           & \lstinline[language=C++]!true!                                               \\
    \lstinline[language=C++]!invariant_mass_cut!     & $m_t-{}$\lstinline[language=C++]!mW!                                      & N/A                                                                          \\
    \lstinline[language=C++]!ml!                     & \lstinline[language=C++]!0!                                               & \lstinline[language=C++]!0!                                                  \\
    \lstinline[language=C++]!r4!                     & \lstinline[language=C++]!824.12! for \lstinline[language=C++]!nl_top = 5! & \lstinline[language=C++]!1220.3! for \lstinline[language=C++]!nl_bottom = 4! \\
    \lstinline[language=C++]!alpha!                  & \lstinline[language=C++]!alpha_mZ = 1/128.944!                            & \lstinline[language=C++]!alpha_Y = 1/132.274!                                \\
    \lstinline[language=C++]!mu_alpha!               & \lstinline[language=C++]!mZ = 91.1876!                                    & \lstinline[language=C++]!mu_alpha_Y = 10.2!                                  \\
    \lstinline[language=C++]!resum_poles!            & \lstinline[language=C++]!6!                                               & \lstinline[language=C++]!6!                                                  \\
    \lstinline[language=C++]!beyond_QCD!             & \lstinline[language=C++]!SM::all!                                         & \lstinline[language=C++]!SM::QED!                                            \\
    \lstinline[language=C++]!mass_scheme!            & \lstinline[language=C++]!{PSshift, mu_f_top}!                             & \lstinline[language=C++]!{PSshift, mu_f_bottom}!                             \\
                                                     & with \lstinline[language=C++]!mu_f_top = 20.!                             & with \lstinline[language=C++]!mu_f_bottom = 2.!                              \\
    \lstinline[language=C++]!production!             & \lstinline[language=C++]!production_channel::all!                         & \lstinline[language=C++]!production_channel::all!                            \\
    \lstinline[language=C++]!expand_s!               & \lstinline[language=C++]!true!                                            & \lstinline[language=C++]!false!                                              \\
    \lstinline[language=C++]!double_light_insertion! & \lstinline[language=C++]!false!                                           & \lstinline[language=C++]!false!                                              \\
    \bottomrule
  \end{tabular}
  \caption{Default option settings in version 1.0 for top and bottom
related functions. Variables in the second and third column are defined
in the header constants.hpp and can be adjusted during or before
installation. The default value for \lstinline[basicstyle=\captioncode,language=C++]!r4! is
adjusted automatically if the number of light flavours is changed in
constants.hpp.}
  \label{tab:defaults}
\end{table}

\section{Advanced usage}
\label{sec:advanced_usage}

In the following we discuss several more complicated examples that
require some knowledge of the options discussed in
section~\ref{sec:options} and the structure of the cross section
outlined in section~\ref{sec:xs_struct}.

\subsection{Wave function at the origin}
\label{sec:wave_function}

While the quarkonium wave function at the origin is not a physical
observable, it serves as a good example to illustrate some of the more
advanced options. Our starting point is eq.~(\ref{eq:ZN_def}), the
definition of the residue $Z_N$. The following assumes that we wish to
determine the wave function at the origin in the PS-shift scheme and
that the \lstinline[language=C++]!expand_s! option is set to false
(which is the default for bottom quarks). In a first step, we eliminate
$d_v$ and the higher-order corrections to $c_v$ with the
\lstinline[language=C++]!contributions! option. We then cancel the
remaining prefactor by multiplying with $s_N/(4m_Q^2)$, where $m_Q$ is
the (order-dependent) pole mass computed from $m_Q^{\text{PS}}$ according to
the shift scheme prescription. To this end, we
need to calculate the binding energy $E_N^{\text{PS}}$ and convert the
input mass from the PS scheme to the pole scheme with the
\lstinline[language=C++]!bottom_pole_mass! function presented in
section~\ref{sec:scheme_conv}. The following code computes
$|\psi_1(0)|^2 = 0.721131\,\text{GeV}^3$ for the $\Upsilon(1S)$
resonance:
\begin{lstlisting}[language=C++]
#include <iostream>

#include "QQbar_threshold/scheme_conversion.hpp"
#include "QQbar_threshold/energy_levels.hpp"
#include "QQbar_threshold/residues.hpp"

int main(){
  namespace QQt = QQbar_threshold;
  const double mb_PS = 4.5;
  const double mu = 4.5;
  QQt::options opt = QQt::bottom_options();
  opt.contributions.cv = {0., 0., 0.};
  opt.contributions.dv = {0., 0.};
  const double E_1_PS = QQt::bbbar_energy_level(
      1, mu, mb_PS, QQt::N3LO, opt
  );
  const double s_1 = (2*mb_PS + E_1_PS)*(2*mb_PS + E_1_PS);
  const double mb_pole = QQt::bottom_pole_mass(
      mu, mb_PS, QQt::N3LO, opt
  );
  std::cout
    << s_1/(4*mb_pole*mb_pole)*QQt::bbbar_residue(
        1, mu, mb_PS, QQt::N3LO, opt
    )
    << '\n';
}
\end{lstlisting}
\begin{lstlisting}[language=Mathematica]
Needs["QQbarThreshold`"];

With[
   {
      mbPS = 4.5,
      mu = 4.5,
      opt = Contributions -> ExceptContributions[
         cv -> {0, 0, 0},
         dv -> {0, 0}
      ]
   },
   E1PS = BBbarEnergyLevel[1, mu, mbPS, "N3LO", opt];
   s1 = (2*mbPS + E1PS)^2;
   mbPole = BottomPoleMass[mu, mbPS, "N3LO", opt];
   Print[s1/(4*mbPole^2)*BBbarResidue[1, mu, mbPS, "N3LO", opt]];
];
\end{lstlisting}
Since $s_N=(2\*m_Q + E_N)^2 = (2\*m_Q^{\text{PS}} + E_N^{\text{PS}})^2$
is a scheme-independent quantity we could also have computed the binding
energy in the pole scheme and combined it with the pole mass. This
method works analogously in the other shift schemes. In an insertion
scheme, we proceed the same way.  However, in this case one needs to
multiply by $s_N/(4\*(m_Q^{\text{RS}})^2)$.

\subsection{Parallelisation}
\label{sec:parallel}

While the observables for a single given set of parameters can be
calculated rather quickly, computing e.g. the cross section over a
range of centre-of-mass energies and parameter settings can become
somewhat time consuming. In such a situation parallelisation can lead to
significant speed-ups.

As an example, we show how to perform a threshold scan similar to
the one discussed in section~\ref{sec:xs}, but including scale
variation. Our strategy is to generate a list containing the
centre-of-mass energy, the cross section obtained with a default scale
of $80\,$GeV, the minimum and the maximum cross section obtained through
scale variation for each point in parallel. Since the mechanisms
typically used in C++ vs. Mathematica programs are rather different, we
discuss these languages separately in the following sections.

\subsubsection{C++}
\label{sec:parallel_C++}

Since we use threads for parallelisation it may be necessary to add
additional flags for compilation. The \texttt{g++} compiler for instance
requires the \texttt{-pthread} option:
\begin{lstlisting}[language=sh]
g++ -o parallel -std=c++11 parallel.cpp -pthread -lQQbar_threshold
\end{lstlisting}
As in the previous examples we use an abbreviation for the somewhat
unwieldy QQbar\_threshold namespace:
\begin{lstlisting}[language=C++]
namespace QQt = QQbar_threshold;
\end{lstlisting}
First, we set up a struct comprising the minimum, maximum, and default
cross sections obtained for a given centre-of-mass energy:
\begin{lstlisting}[language=C++]
struct xs_point{
  double sqrt_s;
  double xs_default;
  double xs_min;
  double xs_max;
};
\end{lstlisting}
Since we call the cross section function many times with mostly the
same arguments, it is also useful to define an auxiliary function
that has only the energy and the scale as remaining arguments:
\begin{lstlisting}[language=C++]
double xsection(double sqrt_s, double mu){
  static constexpr double mu_width = 350.;
  static constexpr double mt = 171.5;
  static constexpr double width = 1.33;
  return QQt::ttbar_xsection(
      sqrt_s,
      {mu, mu_width},
      {mt, width},
      QQt::N3LO
  );
}
\end{lstlisting}
Using this function, we can then define the scale variation for a single
centre-of-mass energy. For the sake of simplicity, we use a rather
crude sampling of the cross section to estimate the extrema.
\begin{lstlisting}[language=C++]
xs_point xsection_scale_variation(double sqrt_s){
  static constexpr double mu_default = 80.;
  static constexpr double mu_min = 50.;
  static constexpr double mu_max = 350.;
  static constexpr double mu_step = 5.;
  xs_point result;
  result.sqrt_s = sqrt_s;
  result.xs_default = xsection(sqrt_s, mu_default);
  result.xs_min = result.xs_default;
  result.xs_max = result.xs_default;
  for(double mu = mu_min; mu < mu_max; mu += mu_step){
    double current_xsection = xsection(sqrt_s, mu);
    if(current_xsection < result.xs_min){
      result.xs_min = current_xsection;
    }
    else if(current_xsection > result.xs_max){
      result.xs_max = current_xsection;
    }
  }
  return result;
}
\end{lstlisting}
The code to actually calculate the scale variation for various energies
in parallel is then rather short:
\begin{lstlisting}[language=C++]
std::vector<std::future<xs_point>> results;
for(double sqrt_s = 340.; sqrt_s < 349.; sqrt_s += 0.2){
  results.emplace_back(
      std::async(
          std::launch::async,
          xsection_scale_variation, sqrt_s
      )
  );
}
\end{lstlisting}
For each energy, a new thread is launched for computing the scale
variation. While this can be rather inefficient in practice, it still
ilustrates how parallelisation can be achieved in principle.

To complete the example, we should also include the appropriate headers,
load a grid, and produce some output. While all these steps are
straightforward, some care should be taken when loading the grid. As
already stated in section~\ref{sec:xs}, the
\lstinline[language=C++]!load_grid! function must not be called
concurrently from more than one thread. All other functions provided by
\texttt{QQbar\_threshold} can, however, be safely used in a multithreaded
environment. There is one additional exception: if the option
\lstinline[language=C++]!double_light_insertion! is set to
\lstinline[language=C++]!true!, the functions for energy levels and
residues must not be invoked explicitly from different threads at the
same time.\footnote{As already mentioned in section~\ref{sec:options},
for the cross section including pole resummation the
\lstinline[basicstyle=\captioncode,language=C++]!double_light_insertion! option is ignored, so cross
section calculations are always thread safe.}

Finally, here is the full code for the parallelised threshold scan with
scale variation:
\begin{lstlisting}[language=C++]
#include <iostream>
#include <iomanip>
#include <vector>
#include <thread>
#include <future>

#include "QQbar_threshold/load_grid.hpp"
#include "QQbar_threshold/xsection.hpp"

namespace QQt = QQbar_threshold;

struct xs_point{
  double sqrt_s;
  double xs_default;
  double xs_min;
  double xs_max;
};

double xsection(double sqrt_s, double mu){
  static constexpr double mu_width = 350.;
  static constexpr double mt = 171.5;
  static constexpr double width = 1.33;
  return QQt::ttbar_xsection(
      sqrt_s,
      {mu, mu_width},
      {mt, width},
      QQt::N3LO
  );
}

xs_point xsection_scale_variation(double sqrt_s){
  static constexpr double mu_default = 80.;
  static constexpr double mu_min = 50.;
  static constexpr double mu_max = 350.;
  static constexpr double mu_step = 5.;
  xs_point result;
  result.sqrt_s = sqrt_s;
  result.xs_default = xsection(sqrt_s, mu_default);
  result.xs_min = result.xs_default;
  result.xs_max = result.xs_default;
  for(double mu = mu_min; mu < mu_max; mu += mu_step){
    double current_xsection = xsection(sqrt_s, mu);
    if(current_xsection < result.xs_min){
      result.xs_min = current_xsection;
    }
    else if(current_xsection > result.xs_max){
      result.xs_max = current_xsection;
    }
  }
  return result;
}

int main(){
  QQt::load_grid(QQt::grid_directory() + "ttbar_grid.tsv");
  std::vector<std::future<xs_point>> results;
  for(double sqrt_s = 340.; sqrt_s < 349.; sqrt_s += 0.2){
    results.emplace_back(
        std::async(
            std::launch::async,
            xsection_scale_variation, sqrt_s
        )
    );
  }
  std::cout << std::fixed;
  std::cout << "sqrt_s    \tcentral \tmin     \tmax\n";
  for(auto & res: results){
    xs_point current_point = res.get();
    std::cout << current_point.sqrt_s << '\t'
              << current_point.xs_default << '\t'
              << current_point.xs_min << '\t'
              << current_point.xs_max << '\n';
  }
}
\end{lstlisting}

\subsubsection{Mathematica}
\label{sec:parallel_Math}

For parallelisation in Mathematica, we have to ensure that each kernel
knows all relevant definitions and has its own precomputed grid:
\begin{lstlisting}[language=Mathematica]
Needs["QQbarThreshold`"];
LaunchKernels[];

ParallelEvaluate[Needs["QQbarThreshold`"]];
ParallelEvaluate[LoadGrid[GridDirectory <> "ttbar_grid.tsv"]];
\end{lstlisting}
This is very different from the parallelisation at thread level we used
in the C++ case. In the present example, the kernels are independent
processes and therefore unable to share a common grid. It is not only
safe, but even necessary to load multiple copies of a grid at the same
time.

For the actual threshold scan, we first define an auxiliary function for
the cross section with fixed top quark properties and width scale:
\begin{lstlisting}[language=Mathematica]
XSection[sqrts_, mu_] := With[
   {muWidth = 350, mt = 171.5, width = 1.33},
   TTbarXSection[sqrts, {mu, muWidth}, {mt, width}, "N3LO"]
];
\end{lstlisting}
To perform the scale variation, we use Mathematica's built-in
\lstinline[language=Mathematica]!NMinValue! and
\lstinline[language=Mathematica]!NMaxValue! functions:
\begin{lstlisting}[language=Mathematica]
XSectionScaleVariation[sqrts_] := Module[
   {
      muDefault = 80, muMin = 50, muMax = 350,
      xsDefault, xsMin, xsMax, mu
   },
   xsDefault = XSection[sqrts, muDefault];
   xsMin = NMinValue[
      {XSection[sqrts, mu], muMin <= mu <= muMax},
      mu,
      Method -> "SimulatedAnnealing"
   ];
   xsMax = NMaxValue[
      {XSection[sqrts, mu], muMin <= mu <= muMax},
      mu,
      Method -> "SimulatedAnnealing"
   ];
   Return[{sqrts, xsDefault, xsMin, xsMax}];
];
\end{lstlisting}
After this, the code for the actual parallelised scan is again rather
compact:
\begin{lstlisting}[language=Mathematica]
results = ParallelTable[
   XSectionScaleVariation[sqrts],
   {sqrts, 340, 349, 0.2}
];
\end{lstlisting}
We complete the example by adding some code for the output:
\begin{lstlisting}[language=Mathematica]
LaunchKernels[];

Needs["QQbarThreshold`"];
ParallelEvaluate[Needs["QQbarThreshold`"]];
ParallelEvaluate[LoadGrid[GridDirectory <> "ttbar_grid.tsv"]];

XSection[sqrts_, mu_] := With[
   {muWidth = 350, mt = 171.5, width = 1.33},
   TTbarXSection[sqrts, {mu, muWidth}, {mt, width}, "N3LO"]
];

XSectionScaleVariation[sqrts_] := Module[
   {
      muDefault = 80, muMin = 50, muMax = 350,
      xsDefault, xsMin, xsMax, mu
   },
   xsDefault = XSection[sqrts, muDefault];
   xsMin = NMinValue[
      {XSection[sqrts, mu], muMin <= mu <= muMax},
      mu,
      Method -> "SimulatedAnnealing"
   ];
   xsMax = NMaxValue[
      {XSection[sqrts, mu], muMin <= mu <= muMax},
      mu,
      Method -> "SimulatedAnnealing"
   ];
   Return[{sqrts, xsDefault, xsMin, xsMax}];
];

results = ParallelTable[
   XSectionScaleVariation[sqrts],
   {sqrts, 340, 349, 0.2}
];

PrependTo[results, {"sqrt_s", "central", "min", "max"}];
Print[TableForm[results]];
\end{lstlisting}

\subsection{Moments for nonrelativistic sum rules}
\label{sec:Moments}

Next to threshold scans for $t\bar{t}$ production, the calculation of
moments for $\Upsilon$ sum rules is one of the key applications for
\texttt{QQbar\_threshold}. It is conventional to consider the normalised
cross section (cf. eq.~(\ref{eq:sigma_noQED})) $R_b =
\sigma_b/\sigma_{pt}$ with $\sigma_{pt} = 4\*\pi\*\alpha(\mu_\alpha)^2/(3\*s)$,
which can be calculated with the \lstinline[language=C++]!bbbar_R_ratio! function. The moments
of $R_b$ are then defined as
\begin{equation}
  \label{eq:M_def}
  {\cal M}_n = \int_0^\infty \frac{R_b(s)}{s^{n+1}}\,.
\end{equation}
Splitting the moments into the contribution from the narrow $\Upsilon$
resonances and the remaining continuum contribution we obtain
\begin{equation}
  \label{eq:M_res_cont}
  {\cal M}_n = \frac{12 \pi^2 N_c e_b^2}{m_b^2} \sum_{N=1}^\infty
  \frac{Z_N}{s_N^{2\*n+1}} + \int_{4\*m_b^2}^\infty ds\,\frac{R_b(s)}{s^{n+1}}\,.
\end{equation}
We now show how this formula can be evaluated with
\texttt{QQbar\_threshold}. Since discussing
numerical integration is clearly outside the scope of this work, we
assume the existence of a C++ header integral.hpp that provides a
suitable \lstinline[language=C++]!integral! function.\footnote{For the
Mathematica corresponding code, we can and do of course use the
built-in \lstinline[basicstyle=\captioncode,language=Mathematica]!NIntegrate! function.} The
example code moments.cpp distributed together with
\texttt{QQbar\_threshold} in fact includes such a header. Since this
header uses the \texttt{GSL} library, the code example has to be
compiled with additional linker flags, e.g.
\begin{lstlisting}[language=sh]
g++ -o moments -std=c++11 moments.cpp -lQQbar_threshold -lgsl -lgslcblas
\end{lstlisting}

For the sake of simplicity, we use the standard bottom options and
only keep the bottom quark mass and $n$ as free parameters in our
example. We can then define auxiliary functions for the energy levels,
residues, and the continuum cross section:
\begin{lstlisting}[language=C++]
static constexpr double pi = 3.14159265358979323846264338328;

double Z(int N, double mb_PS){
  return QQt::bbbar_residue(N, mb_PS, mb_PS, QQt::N3LO);
}

double E(int N, double mb_PS){
  return QQt::bbbar_energy_level(N, mb_PS, mb_PS, QQt::N3LO);
}

double Rb(double s, double mb_PS){
  try{
    return QQt::bbbar_R_ratio(std::sqrt(s), mb_PS, mb_PS, QQt::N3LO);
  }
  catch(std::out_of_range){
    return std::numeric_limits<double>::quiet_NaN();
  }
}
\end{lstlisting}
Note that for the bottom cross section we have to handle the case that
the centre-of-mass energy is outside the region covered by the
precomputed grid. In fact, we assume that the value obtained for
the continuum integral in eq.~(\ref{eq:M_res_cont}) is reliable in spite
of the limited grid size. For a more careful analysis, this assumption
should of course be checked, for example by varying the upper bound of
the integral.

Let us first consider the resonance contribution. For the prefactor, we
have to convert the input mass from the PS scheme to the pole scheme,
which can be done with the function
\lstinline[language=C++]!bottom_pole_mass! (see section~\ref{sec:scheme_conv}). The
remaining code is then rather straightforward:
\begin{lstlisting}[language=C++]
double M_resonances(int n, double mb_PS){
  static constexpr int N_c = 3;
  static constexpr double e_b = QQt::e_d;

  double sum_N = 0.0;
  for(int N = 1; N <= 6; ++ N){
    sum_N += Z(N, mb_PS)*std::pow(2*mb_PS + E(N, mb_PS), -2*n-1);
  }
  const double mb_pole = QQt::bottom_pole_mass(mb_PS, mb_PS, QQt::N3LO);
  return 12*pi*pi*N_c*e_b*e_b/(mb_pole*mb_pole)*sum_N;
}
\end{lstlisting}
Since at N${}^3$LO, only the first six resonances can be computed with
\texttt{QQbar\_threshold}, we have cut off the sum at $N=6$.

For the continuum contribution, we encounter the problem that typical
C++ integration routines can only evaluate integrals over a finite
interval. To deal with this we perform a substitution, e.g. $s =
s(x) = 4\*m_b^2 + \frac{x}{1-x}$, so we have
\begin{equation}
  \label{eq:M_cont_subst}
  \int_{4\*m_b^2}^\infty ds\, \frac{R_b(s)}{s^{n+1}} = \int_0^1
  \frac{dx}{(1 - x)^2} \frac{R_b\big(s(x)\big)}{s(x)^{n+1}}\,.
\end{equation}
The continuum moments can then be computed with the following code:
\begin{lstlisting}[language=C++]
double M_continuum(int n, double mb_PS){
  double const s0 =
    pow(QQt::bbbar_threshold(mb_PS, mb_PS, QQt::N3LO), 2);
  auto s = [=](double x){
    return s0 + x/(1-x);
  };
  auto integrand = [=](double x){
    return Rb(s(x), mb_PS)*std::pow(s(x), -n-1)*std::pow(1 - x, -2);
  };
  return integral(0, 1, integrand);
}
\end{lstlisting}
All that remains is then to add up both contributions and add the
standard boilerplate code for including headers, loading the grid,
etc. It is also convenient to rescale the moments to be of order
one. For this, we multiply them by a factor of
$(10\,\text{GeV})^{2\*n}$. Finally, here is the complete code for our
example:
\begin{lstlisting}[language=C++]
#include <cmath>
#include <limits>
#include <iostream>

#include "QQbar_threshold/load_grid.hpp"
#include "QQbar_threshold/xsection.hpp"
#include "QQbar_threshold/threshold.hpp"
#include "QQbar_threshold/residues.hpp"
#include "QQbar_threshold/energy_levels.hpp"
#include "QQbar_threshold/scheme_conversion.hpp"

#include "integral.hpp"

namespace QQt = QQbar_threshold;

static constexpr double pi = 3.14159265358979323846264338328;

double Z(int N, double mb_PS){
  return QQt::bbbar_residue(N, mb_PS, mb_PS, QQt::N3LO);
}

double E(int N, double mb_PS){
  return QQt::bbbar_energy_level(N, mb_PS, mb_PS, QQt::N3LO);
}

double Rb(double s, double mb_PS){
  try{
    return QQt::bbbar_R_ratio(std::sqrt(s), mb_PS, mb_PS, QQt::N3LO);
  }
  catch(std::out_of_range){
    return std::numeric_limits<double>::quiet_NaN();
  }
}

double M_resonances(int n, double mb_PS){
  static constexpr int N_c = 3;
  static constexpr double e_b = QQt::e_d;

  double sum_N = 0.0;
  for(int N = 1; N <= 6; ++ N){
    sum_N += Z(N, mb_PS)*std::pow(2*mb_PS + E(N, mb_PS), -2*n-1);
  }
  const double mb_pole = QQt::bottom_pole_mass(mb_PS, mb_PS, QQt::N3LO);
  return 12*pi*pi*N_c*e_b*e_b/(mb_pole*mb_pole)*sum_N;
}

double M_continuum(int n, double mb_PS){
  double const s0 =
    pow(QQt::bbbar_threshold(mb_PS, mb_PS, QQt::N3LO), 2);
  auto s = [=](double x){
    return s0 + x/(1-x);
  };
  auto integrand = [=](double x){
    return Rb(s(x), mb_PS)*std::pow(s(x), -n-1)*std::pow(1 - x, -2);
  };
  return integral(0, 1, integrand);
}

double M(int n, double mb_PS){
  return pow(10, 2*n)*(M_resonances(n, mb_PS) + M_continuum(n, mb_PS));
}

int main(){
  QQt::load_grid(QQt::grid_directory() + "bbbar_grid_large.tsv");
  std::cout << M(10, 4.5322) << '\n';
}
\end{lstlisting}
In units of $(10\,\text{GeV})^{-20}$, we find ${\cal M}_{10} =
0.264758$, which is in good agreement with the experimental value ${\cal
M}_{10} = 0.2648(36)$. In other words, our determination of the PS mass
agrees very well with the central value $m_b^{\text{PS}} = 4.532\,$GeV
found
in~\cite{Beneke:2014pta}.\footnote{In~\cite{Beneke:2014pta} QED effects
were treated differently and a non-zero mass was used for the light
quark. The numerical effect on the extracted PS mass is negligible.}

For Mathematica, the structure is quite similar. To avoid clashes with
built-in functions, we have renamed some of the variables.
\begin{lstlisting}[language=Mathematica]
Needs["QQbarThreshold`"];

ZN[i_, mbPS_] := BBbarResidue[i, mbPS, mbPS, "N3LO"];

EN[i_, mbPS_] := BBbarEnergyLevel[i, mbPS, mbPS, "N3LO"];

Rb[s_, mbPS_] := BBbarRRatio[Sqrt[s], mbPS, mbPS, "N3LO"];

Mresonances[n_, mbPS_] := With[
   {
      Nc = 3, eb = eD,
      mbPole = BottomPoleMass[mbPS, mbPS, "N3LO"]
   },
   12*Pi^2*Nc*eb^2/mbPole^2*Sum[
      ZN[i, mbPS]/(2*mbPS + EN[i, mbPS])^(2*n+1),
      {i, 1, 6}
   ]
];

Mcontinuum[n_, mbPS_] := Module[
   {s0, s, x},
   s0 = BBbarThreshold[mbPS, mbPS, "N3LO"]^2;
   s[x_] := s0 + x/(1 - x);
   Return[
      NIntegrate[
         Rb[s[x], mbPS]/(s[x]^(n+1)*(1 - x)^2),
         {x, 0, 1},
         AccuracyGoal -> 5
      ]
   ];
];

M[n_, mbPS_] := 10^(2*n)*(Mresonances[n, mbPS] + Mcontinuum[n, mbPS]);

LoadGrid[GridDirectory <> "bbbar_grid_large.tsv"];
Print[M[10, 4.5322]];
\end{lstlisting}
Here, we obtain a slightly different value of ${\cal M}_{10} =
0.264857$, which stems from a different integration algorithm. Since
this change can be offset by changing the input mass by a small amount
of about $0.1\,$MeV, we can conclude that integration errors are under
control.

\section{Grid generation}
\label{sec:grid_gen}

Depending on the chosen input parameters, the precomputed grids
distributed with \texttt{QQbar\_threshold} may not be sufficient. In
these cases custom grids can be generated with the
\texttt{QQbarGridCalc} package, which can be downloaded separately from
\url{https://www.hepforge.org/downloads/qqbarthreshold/} and needs no
special installation aside from unpacking the archive with
\lstinline[language=bash]!tar xzf QQbarGridCalc.tar.gz!. Grid generation
requires Wolfram Mathematica. The Mathematica working directory should
be set to the \texttt{QQbarGridCalc} directory (e.g. with the
\lstinline[language=Mathematica]!SetDirectory! command), so that all
included files are found.

\subsection{Top and bottom grids}
\label{sec:top_bottom_grids}

The main function provided by this
package is \lstinline[language=Mathematica]!QQbarCalcGrid!, which
generates a grid for the bottom or top cross section. It can be used in the
following way:
\begin{lstlisting}[language=Mathematica]
QQbarCalcGrid[
   Energy -> {MinEnergy, MaxEnergy, EnergyStep},
   Width -> {MinWidth, MaxWidth, WidthStep},
   "GridFileName"
];
\end{lstlisting}
\texttt{MinEnergy} and \texttt{MaxEnergy} refer to the na\"ive threshold
at $\sqrt{s} = 2m_Q$, where $m_Q$ is the heavy-quark pole mass. The
generated grid thus covers the
centre-of-mass energies $2m_Q + \mathtt{MinEnergy} \leq \sqrt{s} \leq
2m_Q + \mathtt{MaxEnergy}$ and the widths $\mathtt{MinWidth} \leq \Gamma \leq
\mathtt{MaxWidth}$. \texttt{EnergyStep} and \texttt{WidthStep}
specify the distance between adjacent grid points. The resulting grid
is saved in the file \texttt{GridFileName}. For example, the following
program creates a small top grid and exports it to the file \texttt{top\_grid\_example.tsv}:
\begin{lstlisting}[language=Mathematica]
<<QQbarGridCalc.m;

LaunchKernels[];

QQbarCalcGrid[
   Energy -> {-1, 1, 1},
   Width -> {1.5, 1.6, 0.1},
   "top_grid_example.tsv"
];
\end{lstlisting}
Note that loading the package typically takes several minutes. The
calculation of the grids themselves is even more time-consuming, so we
restrict the examples to very small and coarse grids and suggest to
rely on parallelisation as much as possible.

The energy and width ranges always refer to reference values for the
quark mass and the strong coupling, specified with the
\lstinline[language=Mathematica]!QuarkMass! and
\lstinline[language=Mathematica]!AlphaS! options (defaulting to 175 and
0.14, respectively). In fact, internally all energies and widths are rescaled
by a factor of $-m_Q\*\alpha_s^2\*C_F^2/4$. In practice, this implies that the range
covered in the actual calculation of the cross section will in general
be slightly different. Furthermore, the default values for
\lstinline[language=Mathematica]!QuarkMass! and
\lstinline[language=Mathematica]!AlphaS! are chosen with top grids in
mind, so one should change these settings when calculating bottom grids.

In some cases it is desirable to have grids that are relatively coarse
in one region, e.g. at high energies, and much finer in another
region. To this end it is possible to directly specify the energy and
width points when calling
\lstinline[language=Mathematica]!QQbarCalcGrid! as
\begin{lstlisting}[language=Mathematica]
  QQbarCalcGrid[
   Energy -> {{EnergiesPts ... }},
   Width -> {{WidthPts ... }},
   "GridFileName"
];
\end{lstlisting}
The following example
shows how a bottom grid with a higher resolution close to the threshold
can be generated:
\begin{lstlisting}[language=Mathematica]
<<QQbarGridCalc.m;

LaunchKernels[];

emin = 10^-6;
emax = 10;
n = 3;
epoints = Module[
   {stepfact},
   stepfact = (emax/emin)^(1/(n - 1));
   Table[N[emin*stepfact^i], {i, 0, n - 1}]
];

QQbarCalcGrid[
   Energy -> {epoints},
   Width -> {{10^-10, 10^-8}},
   "bottom_grid_example.tsv",
   QuarkMass -> 5,
   AlphaS -> 0.25
];
\end{lstlisting}
Note that the numerical evaluation requires at least a small
non-vanishing width, which is internally set to $10^{-9}$ for bottom
quarks. For such a small width it is not possible (and not very useful)
to calculate grid points with negative energies.

Finally, \lstinline[language=Mathematica]!QQbarCalcGrid! offers the
\lstinline[language=Mathematica]!Comments! option to prepend custom
comments to the generated grid file:
\begin{lstlisting}[language=Mathematica]
QQbarCalcGrid[
   Energy -> {...},
   Width -> {...},
   "GridFileName",
   Comments -> {
     "Comment in the first line of the grid file",
     "Comment in the second line of the grid file"
   }
];
\end{lstlisting}
The default setting
\lstinline[language=Mathematica]!Comment -> Automatic!
adds the version of \lstinline[language=Mathematica]!QQbarCalcGrid!, a
shortened version of the command used for the creation, and the creation
date.

\subsection{Nonresonant grids}
\label{sec:nr_grids}

The second function in \texttt{QQbarGridCalc},
\lstinline[language=Mathematica]!QQbarCalcNonresonantGrid!, allows the
generation of grids for the nonresonant cross section (see
section~\ref{sec:nr_xs}). Its syntax is similar to
\lstinline[language=Mathematica]!QQbarCalcGrid!:
\begin{lstlisting}[language=Mathematica]
  QQbarCalcNonresonantGrid[
   MassRatio -> {...},
   Cut -> {...},
   "GridFileName"
];
\end{lstlisting}
As with \lstinline[language=Mathematica]!QQbarCalcGrid! both regular and
irregular grids can be generated and also the
\lstinline[language=Mathematica]!Comment! option is supported.  The
first argument specifies the mass ratios $x = m_W/m_Q$, whereas the
second argument determines the invariant mass cut. The coordinates
entered here correspond to $y_w = (1 - y)/(1 - x)$, where $y = (1 -
\Delta_m/m_Q)^2$ and $\Delta_m$ is the cut specified by the
\lstinline[language=C++]!invariant_mass_cut! option (see
section~\ref{sec:options}). Thus, for physical cuts $0 \leq y_w \leq
1$. The default built-in nonresonant grid can be reproduced with the
following program:
\begin{lstlisting}[language=Mathematica]
<<QQbarGridCalc.m;

LaunchKernels[];

QQbarCalcNonresonantGrid[
   MassRatio -> {0.15, 0.30, 0.01},
   Cut -> {0, 1, 0.01},
   "non-resonant_grid.tsv"
];
\end{lstlisting}

\section*{Acknowledgements}

We thank K.~Schuller for contributing to an earlier program for
heavy-quark production near threshold, T.~Rauh for cross-checking parts
of the current implementation, and F.~Simon for valuable comments on the
program and the manuscript. We are grateful to the authors
of~\cite{Beneke:2010mp} for the permission to use their code for the
non-resonant cross section.

Y.~K., A.~M., and J.~P. thank the Technische Universit\"at M\"unchen and
the Excellence Cluster ``Origin and Structure of the Universe'' for
hospitality and travel support. A.~M. is grateful to the Mainz Institute
for Theoretical Physics (MITP) for its hospitality and its partial
support during the completion of this work. A.~M. is supported by a
European Union COFUND/Durham Junior Research Fellowship under EU grant
agreement number 267209. The work of Y.~K. was supported in part by
Grant-in-Aid for scientific research Nos. 26400255 from MEXT,
Japan. This work is further supported by the Gottfried Wilhelm Leibniz
programme of the Deutsche Forschungsgemeinschaft (DFG) and the
Excellence Cluster ``Origin and Structure of the Universe'' at
Technische Universit\"at M\"unchen.

\appendix
\renewcommand*{\thesection}{\Alph{section}}

\section{Predefined constants}
\label{sec:const}

Table~\ref{tab:const} lists all predefined constants and their
values. The values can be adjusted prior to or during the installation of
the \texttt{QQbar\_threshold} library.
\begin{table}
  \centering
  \small
  \begin{tabularx}{\textwidth}{lllX}
    \toprule
    C++ name                                        & Mathematica name                                  & Value                               & Description                                                                 \\
    \midrule
    $\dagger$\,\lstinline[language=C++]!alpha_s_mZ! & \lstinline[language=Mathematica]!alphaSmZDefault! & $0.1184$                            & Default value for strong coupling at the scale \lstinline[language=C++]!mZ!. \\
    $\dagger$\,\lstinline[language=C++]!alpha_mZ!   & \lstinline[language=Mathematica]!alphamZ!         & $1/128.944$                         & QED coupling at the scale \lstinline[language=C++]!mZ!.                      \\
    $\dagger$\,\lstinline[language=C++]!alpha_Y!    & \lstinline[language=Mathematica]!alphaY!          & $1/132.274$                         & QED coupling at the scale \lstinline[language=C++]!mu_alpha_Y!.              \\
    $\dagger$\,\lstinline[language=C++]!mu_alpha_Y! & \lstinline[language=Mathematica]!muAlphaY!        & $10.2$                              & Typical scale for $\Upsilon$ resonances.                                     \\
    \lstinline[language=C++]!mZ!                    & \lstinline[language=Mathematica]!mZ!              & $91.1876$                           & Mass of the Z boson.                                                         \\
    \lstinline[language=C++]!mW!                    & \lstinline[language=Mathematica]!mW!              & $80.385$                            & Mass of the W boson.                                                         \\
    \lstinline[language=C++]!G_F!                   & \lstinline[language=Mathematica]!GF!              & $1.1663787 \times 10^{-5}$          & Fermi constant. Only used for calculating the top width.                     \\
    $\dagger$\,\lstinline[language=C++]!m_Higgs!    & \lstinline[language=Mathematica]!mHiggsDefault!   & $125$                               & Default value for mass of the Higgs boson.                                   \\
    \lstinline[language=C++]!alpha_QED!             & \lstinline[language=Mathematica]!alphaQED!        & $1/137.035999074$                   & Fine structure constant.                                                     \\
    \lstinline[language=C++]!e_u!                   & \lstinline[language=Mathematica]!eU!              & $2/3$                               & Electric charge of the top quark in units of the positron charge.            \\
    \lstinline[language=C++]!e_d!                   & \lstinline[language=Mathematica]!eD!              & $-1/3$                              & Electric charge of the bottom quark.                                         \\
    \lstinline[language=C++]!e!                     & \lstinline[language=Mathematica]!eE!              & $-1$                                & Electric charge of the electron.                                             \\
    \lstinline[language=C++]!cw2!                   & \lstinline[language=Mathematica]!cw2!             & \lstinline[language=C++]!mW^2/mZ^2! & Cosine of the weak mixing angle squared.                                     \\
    \lstinline[language=C++]!sw2!                   & \lstinline[language=Mathematica]!sw2!             & \lstinline[language=C++]!1 - cw2!   & Sine of the weak mixing angle squared.                                       \\
    \lstinline[language=C++]!T3_nu!                 & \lstinline[language=Mathematica]!T3Nu!            & $1/2$                               & Weak isospin of neutrino.                                                    \\
    \lstinline[language=C++]!T3_e!                  & \lstinline[language=Mathematica]!T3E!             & $-1/2$                              & Weak isospin of electron.                                                    \\
    \lstinline[language=C++]!T3_u!                  & \lstinline[language=Mathematica]!T3U!             & $1/2$                               & Weak isospin of top.                                                         \\
    \lstinline[language=C++]!T3_d!                  & \lstinline[language=Mathematica]!T3D!             & $-1/2$                              & Weak isospin of bottom.                                                      \\
    \lstinline[language=C++]!mb_SI!                 & \lstinline[language=Mathematica]!mbSI!            & $4.203$                             & Reference scale-invariant mass for bottom quarks.                            \\
    \lstinline[language=C++]!mu_thr!                & \lstinline[language=Mathematica]!muThr!           & \lstinline[language=C++]!2*mb_SI!   & Decoupling threshold for bottom quarks.                                      \\
    \lstinline[language=C++]!nl_bottom!             & \lstinline[language=Mathematica]!nlBottom!        & $4$                                 & Number of light flavours for bottom-related functions.                       \\
    \lstinline[language=C++]!nl_top!                & \lstinline[language=Mathematica]!nlTop!           & $5$                                 & Number of light flavours for top-related functions.                          \\
    $\dagger$ \lstinline[language=C++]!mu_f_bottom! & \lstinline[language=Mathematica]!mufBottom!       & $2$                                 & Default PS scale for bottom.                                                 \\
    $\dagger$ \lstinline[language=C++]!mu_f_top!    & \lstinline[language=Mathematica]!mufTop!          & $20$                                & Default PS scale for top.                                                    \\
    \lstinline[language=C++]!invGeV2_to_pb!         & \lstinline[language=Mathematica]!InvGeV2ToPb!     & $389379300$                         & Conversion factor from GeV$^{-2}$ to picobarn.                               \\
    \bottomrule
  \end{tabularx}
  \caption{Constants predefined in the constants.hpp header. Entries
    marked with a $\dagger$ only serve as default values and can be
    overridden through option settings (cf. section~\ref{sec:options}).}
  \label{tab:const}
\end{table}

\bibliography{biblio}{}
\bibliographystyle{elsarticle-num}

\end{document}